\documentclass{aa}  
\usepackage{graphicx,epsfig,amssymb,dcolumn}
\usepackage{txfonts}
\def\lsim{\vcenter{\hbox{$<$}\offinterlineskip\hbox{$\sim$}}}
\def\gsim{\vcenter{\hbox{$>$}\offinterlineskip\hbox{$\sim$}}}
\newcolumntype{.}{D{.}{.}{4}}
\newcolumntype{,}{D{.}{.}{2}}
\newcolumntype{;}{D{.}{.}{1}}
\begin{document}
%
   \title{Dust, pulsation, chromospheres and their r\^ole in driving mass loss from red giants in Galactic globular clusters}


   \author{I. McDonald
          \inst{1}
          \and
          J. Th. van Loon\inst{1}
          }

   \offprints{I. McDonald}

   \institute{$^1$Astrophysics Group, School of Physical \& Geographical Sciences,
           Keele University, Staffordshire ST5 5BG, UK\\
              \email{iain@astro.keele.ac.uk}
             }

   \date{Received date; accepted date}

 
  \abstract
   {Mass loss from red giants in old globular clusters affects the horizontal branch (HB) morphology and post-HB stellar evolution including the production of ultraviolet-bright stars, dredge up of nucleosynthesis products and replenishment of the intra-cluster medium. Studies of mass loss in globular clusters also allows one to investigate the metallicity dependence of the mass loss from cool, low-mass stars down to very low metallicities.
}
   {We present an analysis of new VLT/UVES spectra of 47 red giants in the Galactic globular clusters 47 Tuc (NGC 104), NGC 362, $\omega$ Cen (NGC 5139), NGC 6388, M54 (NGC 6715) and M15 (NGC 7078). The spectra cover the wavelength region 6100--9900 \AA\ at a resolving power of $R = 110,000$. Some of these stars are known to exhibit mid-infrared excess emission indicative of circumstellar dust. Our aim is to detect signatures of mass loss, identify the mechanism(s) responsible for such outflows, and measure the mass-loss rates.}
   {We determine for each star its effective temperature, luminosity, radius and escape velocity. We analyse the H$\alpha$ and near-infrared calcium triplet lines for evidence of outflows, pulsation and chromospheric activity, and present a simple model for estimating mass-loss rates from the H$\alpha$ line profile. We compare our results with a variety of other, independent methods.}
   {We argue that a chromosphere persists in Galactic globular cluster giants and controls the mass-loss rate to late-K/early-M spectral types, where pulsation becomes strong enough to drive shock waves at luminosities above the RGB tip. This transition may be metallicity-dependent. We find mass-loss rates of $\sim$10$^{-7}$ to $10^{-5}$ M$_{\odot}$ yr$^{-1}$, largely independent of metallicity.}
   {}

   \keywords{
		Stars: AGB and post-AGB --
		Stars: circumstellar matter --
		Stars: mass-loss --
		Stars: chromospheres --
		Infrared: stars --
		Galaxy: globular clusters: individual: NGC 104, NGC 362, NGC 5139, NGC 6388, NGC 6715, NGC 7078
               }

   \titlerunning{Dust, pulsation, chromospheres and mass loss in red giants in Galactic GCs}
   \authorrunning{I. McDonald \& J. Th. van Loon}
   
   \maketitle
%



\section{Introduction}

Understanding the evolution and mass loss in red giant branch (RGB) and asymptotic giant branch (AGB) stars is of substantial importance in uncovering the history of metal enrichment of the inter-stellar medium (ISM) and subsequent stellar and planetary formation; as well as understanding the evolution of the stars themselves. By observing mass-losing giant branch stars in globular clusters (GCs), we can simultaneously observe a co-eval set of objects within a cluster with similar progenitor mass and identical metallicity; while we can use different clusters to observe the effects of differing ages and masses, but more particularly metallicities, which can range from near-solar metallicity to as little as $\sim$0.5\% of solar.

Simultaneously, we also glean information about the globular cluster environment itself -- the mass lost from stars forms the intra-cluster medium (ICM). Studies of the ICM show it is removed from the cluster on a timescale much shorter than the time between Galactic Plane crossings (e.g. Boyer et al.\ 2006; van Loon et al.\ 2006a). This mass loss exacerbates cluster evaporation, possibly leading to their dispersal on a timescale of $10^{9-10}$ years (van Loon \& McDonald 2007).

Both the RGB and AGB are associated with episodes of strong mass loss, eventually leading to stellar death. Globular cluster stars are thought to lose around 0.2 M$_{\odot}$ over their time on the RGB (Rood 1973) with up to 0.1 M$_{\odot}$ being lost in the helium flash at the RGB tip (Fusi-Pecci \& Renzini 1975, and references therein). This is required to explain the horizontal branch (HB) morphology, which is dependent on the remaining mantle mass. Dupree et al.\ (2007) have now observationally confirmed mass loss three magnitudes below the RGB tip, thought to be driven by hydrodynamically- or acoustically-dominated chromospheres.

Stars reaching the upper slopes of the AGB are even more extreme, losing mass at rates reaching well over $10^{-6}$ M$_{\odot}$ yr$^{-1}$ (Wood et al.\ 1983, 1992; van Loon et al.\ 1999). This is facilitated by thermal pulses -- periodically occurring temporary ignition of a helium shell-burning source inside the star -- and shorter-timescale ($\sim$10$^{2-3}$ days) radial pulsations at the surface which can lead to shock-wave emission in the extended stellar atmosphere (Fox \& Wood 1985). In metal-rich AGB stars, dust forming in the outmost parts of these atmospheres is subject to radiation pressure, which forces it and the gas it collides with to be radially accelerated away from the star. It is not yet clear if dust-driving is the dominating effect in RGB and metal-poor AGB winds, and it is possible that Alfv\'en waves or pressure (P-mode) waves dominate here (Hartmann \& MacGregor 1980; Pijpers \& Habing 1989). Indeed, Judge \& Stencel (1991, hereafter JS91) argue that the mass-loss rate is largely invariant with the driving process and that most cool giants (RGB or AGB) do not have dust as the primary driver of their winds. More recently, Schr\"oder \& Cuntz (2005, hereafter SC05) have suggested an improved Reimers relation (Reimers 1975), by assuming this invariance and using simple chromospheric energy considerations, which is calibrated using the total RGB mass-loss of GC stars.

Previous works have attempted to provide evidence for mass loss and calculate mass-loss rates by analysing optical spectral line profiles, either through asymmetries in the line cores (first performed by Deutsch (1956), for the field giant $\alpha$ Her), line emission (Cohen 1976) or modelling. Line emission has also been used to model a chromosphere with a large bulk motion (Dupree et al.\ 1984; Mauas et al.\ 2006). Semi-empirical methods of finding mass-loss rates have also been used to define relationships based on stellar parameters. Mass-loss rates can also be derived from dust masses calculated from IR excess emission (e.g. Origlia et al.\ 2002 -- hereafter OFFR02; van Loon 2007), though this must assume a value for the ratio of the masses of dust and gas lost. Mira-type long-period variability is only seen in stars with [Fe/H] $\gtrsim -1$ (Frogel \& Whitelock 1998), so one would expect that, assuming pulsation assists in driving mass-loss, the mass-loss rate and/or mechanism would vary with metallicity.

In this work, we use all of the above techniques to try to identify mass loss and, where possible, estimate mass-loss rates from a sample of 47 such objects from a range of clusters. Ultimately, we aim to assess the r\^ole of dust, pulsation and chromospheric activity in driving mass loss, and its dependence on metallicity. Our observations are based on new VLT/UVES echelle spectra at very high spectral resolution. We present the spectra and literature photometry in Section 2 of the paper. In Section 3, we use the Kurucz {\sc atlas9} models (Kurucz 1993) to estimate the effective temperature, metallicity and surface gravity of the stars, then use the near-IR photometry to calculate other physical parameters (luminosity, radius and escape velocity). A full abundance analysis has not been performed, but will become the subject of a subsequent paper. In Section 4, we analyse the H$\alpha$ and near-IR Ca II triplet lines and from them provide evidence of outflow from the stellar surface. We present a simple model for estimating mass loss in the stellar wind from the H$\alpha$ profile in Section 5, and use it to calculate mass-loss rates and wind velocities for our sample of stars. This is followed by a discussion in Section 6, in which we also estimate mass-loss rates via other procedures. Our conclusions are presented in Section 7.


\section{Observations and target selection}


\begin{table*}[!htp]
\begin{center}
\caption{List of clusters observed.}
\label{ClusterTable}
\begin{tabular}{l@{\,}.@{}.@{}.@{}.@{}.@{}c@{}c@{}c@{}c}
    \hline \hline
\multicolumn{1}{l}{NGC}&	\multicolumn{1}{c}{[Fe/H]}&	\multicolumn{1}{c}{Heliocentric}&	\multicolumn{1}{c}{Velocity}&	\multicolumn{1}{c}{Distance}&	\multicolumn{1}{c}{Absolute}&	\multicolumn{3}{c}{Stars with(out) IR Excess}&	\multicolumn{1}{c}{Ref-}\\
\multicolumn{1}{l}{Number}&&\multicolumn{1}{c}{Radial Velocity}	&\multicolumn{1}{c}{Dispersion}	&			&\multicolumn{1}{c}{Visual}	
&\multicolumn{3}{c}{$\overbrace{\qquad\qquad\qquad\qquad\qquad}$}
&\multicolumn{1}{c}{erences}\\
&&\multicolumn{1}{c}{(km s$^{-1}$)}&\multicolumn{1}{c}{(km s$^{-1}$)}&\multicolumn{1}{c}{(kpc)}&\multicolumn{1}{c}{Magnitude}
&\multicolumn{1}{c}{With}&\multicolumn{1}{c}{Without}&\multicolumn{1}{c}{Unknown}&\\
    \hline
104 (47\,Tuc)       & -0.76     &   -18.7	& 11.5	&  4.5	&   -9.42 & 9 &  0 & 0 & 1\\
362                 & -1.16     &   223.5	&  7.5	&  8.5	&   -8.41 & 3 & 11 & 2 & 2\\
5139 ($\omega$\,Cen)& -1.62\dag &   232.2	& 19.6	&  5.3	&  -10.29 & 1 &  0 & 0 & 1\\
6388                & -0.60     &    81.2	& 18.9	& 10.0	&   -9.42 & 6 & 12 & 0 & 3\\
6715 (M\,54)        & -1.58     &   141.9	& 14.2	& 26.8	&  -10.01 & 0 &  0 & 1 & 2\\
7078 (M\,15)        & -2.26     &  -107.0	& 11.0	& 10.3	&   -9.17 & 2 &  0 & 0 & 1\\
    \hline
\multicolumn{10}{l}{Metallicity, radial velocity and distance values from Harris (1996).}\\
\multicolumn{10}{l}{Other references: 1 -- Paturel \& Garnier (1992); 2 -- McLaughlin \& van der Marel (2005); 3 -- Gebhardt et al.\ (1997).}\\
\multicolumn{10}{l}{\dag Metal-rich ([Fe/H] $\sim$ --0.6) and metal-intermediate (--1.3 $<$ [Fe/H] $<$ --1.0) populations are also believed to be}\\
\multicolumn{10}{l}{\ \ present (Norris \& da Costa 1995; Pancino et al.\ 2000; Sollima et al.\ 2005).}\\
    \hline
\end{tabular}
\end{center}
\end{table*}

\begin{table*}[!htp]
\begin{center}
\caption{List of objects observed with positions, photometry and ancillary data.}
\label{ObjectTable}
\begin{tabular}{lcc..clr.l}
    \hline \hline
\multicolumn{1}{l}{Our}	& \multicolumn{1}{c}{RA}& \multicolumn{1}{c}{Dec}	& \multicolumn{1}{c}{J}		& \multicolumn{1}{c}{K}		& \multicolumn{1}{c}{IR}	& \multicolumn{1}{c}{Other}	& \multicolumn{1}{c}{Period}	& \multicolumn{1}{c}{Amplitude}	& \multicolumn{1}{l}{Notes}\\
\multicolumn{1}{l}{ID}	& \multicolumn{1}{c}{(J2000.0)}& \multicolumn{1}{c}{(J2000.0)}& \multicolumn{1}{c}{Mag}	& \multicolumn{1}{c}{Mag}	& \multicolumn{1}{c}{Excess?}	& \multicolumn{1}{c}{Names}	& \multicolumn{1}{c}{(days)}	& \multicolumn{1}{c}{(K-band)}	& \\
    \hline 

\multicolumn{10}{c}{NGC\,362} \\
o01 & 01  03  15.1 & --70  50  32 &  9.93 &  8.78 & N &  V16	& 138	& 0.5	& 7\\[-0pt]
o02 & 01  03  13.6 & --70  50  59 & 10.27 &  9.51 & N &  	&	&	&  \\[-0pt]
o03 & 01  03  13.8 & --70  51  09 &  9.60 &  8.79 & N &  \\[-0pt]
o04 & 01  03  12.6 & --70  50  38 & 10.05 &  9.13 & N &  \\[-0pt]
o05a& 01  03  17.1 & --70  50  50 &  9.91:&  9.19:& N & 	&	&	& 2, 5 \\[-0pt]
o05b& 01  03  17.3 & --70  50  50 &  9.95:&  9.27:& N &  	&	&	& 2, 5\\[-0pt]
o06 & 01  03  13.6 & --70  50  37 &  9.68 &  8.74 & N &  	&	& 0.15:	& 7\\[-0pt]
o07 & 01  03  20.1 & --70  50  55 &  9.58 &  8.75 & N &  \\[-0pt]
o08 & 01  03  10.7 & --70  50  54 &  9.85 &  8.89 & N &  \\[-0pt]
o09 & 01  03  10.9 & --70  50  59 & 10.66 &  9.83 & N &  \\[-0pt]
o10 & 01  03  14.7 & --70  51  15 & 10.38 &  9.67 & N &  \\[-0pt]
x01 & 01  03  19.0 & --70  50  52 &  9.96 &  8.96 & Y &  \\[-0pt]
x02a& 01  03  14.7 & --70  50  59 &  9.64:&  8.27:& Y\rlap{?}& 	&	&	& 2 \\[-0pt]
x02b& 01  03  14.9 & --70  51  00 &  9.82:&  8.45:& Y\rlap{?}& 	&	&	& 2, 5 \\[-0pt]
x02c& 01  03  14.5 & --70  50  59 & 10.45 &  9.64 & Y & 	&	&	& 3, 5 \\[-0pt]
x03 & 01  03  13.7 & --70  51  14 & 10.10 &  8.82 & Y &  \\
    \hline
\multicolumn{10}{c}{47\,Tuc} \\
x01 & 00  24  02.6 & --72  05  07 &  7.73 &  6.75 & Y & LW10	& 110	&	& \\[-0pt]
x02 & 00  24  07.8 & --72  05  09 &  8.80 &  7.79 & Y &  \\[-0pt]
x03 & 00  24  09.4 & --72  04  49 &  8.37 &  7.35 & Y &  	&	& 0.15:	& 7\\[-0pt]
x04 & 00  23  57.7 & --72  05  30 &  8.77 &  7.75 & Y & LW8	& 27	& 0.15:	& 7\\[-0pt]
x05 & 00  24  23.2 & --72  04  23 &  8.30 &  7.16 & Y & LW19	& 40	&	&  \\[-0pt]
x06 & 00  24  08.6 & --72  03  55 &  8.09 &  6.84 & Y & V8?	& 155	& 0.3	& 7\\[-0pt]
x07 & 00  24  07.9 & --72  04  32 &  7.82 &  6.56 & Y & LW13	& 65	&	&  \\[-0pt]
x08 & 00  25  16.0 & --72  03  55 &  7.66 &  6.45 & Y & V3	& 192	& 0.6	& 1, 7, 8 \\[-0pt]
x09 & 00  25  09.2 & --72  02  40 &  8.67 &  6.56 & Y & V18	& 83	&	& 1\\
    \hline
\multicolumn{10}{c}{NGC\,6388} \\
o01 & 17  36  14.9 & --44  43  12 & 10.73 &  9.31 & N &  V9\\[-0pt]
o02 & 17  36  11.4 & --44  44  08 & 10.02 &  8.80 & N &  V12	&	& 0.25:	& 7\\[-0pt]
o03 & 17  36  20.2 & --44  44  32 & 11.01 &  9.65 & N &  \\[-0pt]
o04 & 17  36  17.6 & --44  43  46 & 10.49 &  9.29 & N &  \\[-0pt]
o05 & 17  36  16.7 & --44  43  28 & 10.86 &  9.64 & N &  \\[-0pt]
o06 & 17  36  17.7 & --44  44  22 & 10.37 &  9.27 & N & 	&	&	& 1 \\[-0pt]
o07 & 17  36  15.5 & --44  44  26 & 10.72 &  9.31 & N &  \\[-0pt]
o08 & 17  36  15.3 & --44  44  31 & 10.91 &  9.47 & N &  \\[-0pt]
o09 & 17  36  18.9 & --44  44  14 & 11.31 & 10.13 & N &  \\[-0pt]
o10 & 17  36  16.0 & --44  43  49 & 10.73 &  9.72 & N &  \\[-0pt]
o11 & 17  36  20.4 & --44  44  16 & 11.12 &  9.97 & N &  \\[-0pt]
o12 & 17  36  18.9 & --44  44  04 & 11.09 &  9.81 & N &  \\[-0pt]
x01 & 17  36  18.1 & --44  43  25 & 10.80 &  9.75 & Y &  V8?	&	& 0.4	& 7\\[-0pt]
x02 & 17  36  13.1 & --44  43  05 & 10.65 &  9.47 & Y &  \\[-0pt]
x03 & 17  36  15.1 & --44  43  33 & 10.27 &  8.99 & Y &  V3	& 156	& 0.3	& 7\\[-0pt]
x04 & 17  36  21.4 & --44  43  42 & 11.15 &  9.87 & Y &  \\[-0pt]
x05 & 17  36  18.8 & --44  43  46 & 10.60 &  9.25 & Y &  \\[-0pt]
x06 & 17  36  16.0 & --44  44  36 & 10.64 &  9.19 & Y &  \\
    \hline
\multicolumn{10}{c}{M\,15} \\
x01 & 21  29  58.9 &  +12  10  18 & 11.18 & 10.50 & Y & AC761\\[-0pt]
x02 & 21  29  58.2 &  +12  09  47 & 10.63 &  9.92 & Y & AC414	&	&	& 9\\
    \hline
\multicolumn{10}{c}{M\,54} \\
x01 & 18  55  04.5 & --30  29  35 & 11.74 & 11.12 & Y & \\
    \hline
\multicolumn{10}{c}{$\omega$\,Cen} \\
x01 & 13  26  46.4 & --47  29  30 &  9.70 &  8.75 & Y & V42	& 149	& 0.5	& 7, 8\\
    \hline
\multicolumn{10}{l}{\small Notes: Astrometry/photometry from OFFR02 or their unpublished data unless noted otherwise. \normalsize} \\
\multicolumn{10}{l}{\small Photometry sources: (1) 2MASS (corrected to the photometric system of OFFR02); (2) UVES acquisition images \normalsize} \\
\multicolumn{10}{l}{\small \ \ \ \  (see text); (3) VFO04. Colons denote uncertain values. \normalsize} \\
\multicolumn{10}{l}{\small Astrometry sources: (4) 2MASS; (5) UVES acquisition images (see text); (6) VFO04 \normalsize} \\
\multicolumn{10}{l}{\small Variability sources: LW indicates variables from Lebzelter \& Wood (2005). (7) Matsunaga et al.\ (2007), Ita et al.\ (2007) \normalsize} \\
\multicolumn{10}{l}{\small \ \ \ \ or Matsunaga (private communication); (8) Feast et al.\ (2002). \normalsize} \\
\multicolumn{10}{l}{\small (9) identified with 2MASS J21295815+1209466, which is bright at 24$\mu$m (M$_{24} = $1.66 -- Boyer et al.\ 2006). \normalsize} \\
\multicolumn{10}{l}{\small AC identifiers are for (non-variable) sources from Auri\`ere \& Cordoni (1981). \normalsize} \\
    \hline
    \\
    \\
\end{tabular}
\end{center}
\end{table*}

\begin{figure*}[!pth]
\resizebox{\hsize}{!}{\includegraphics{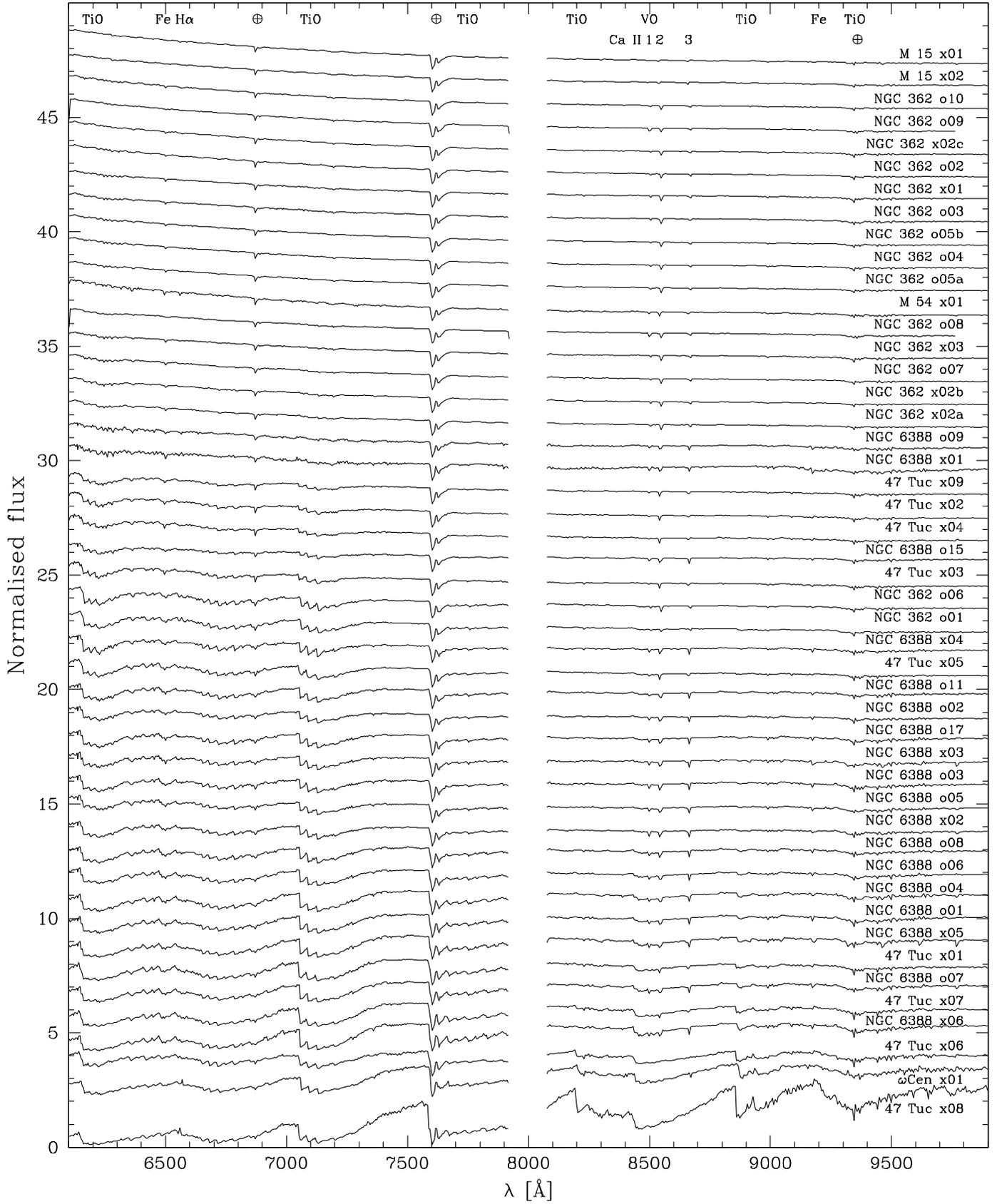}}
\caption[]{Overview of the VLT/UVES spectra of the target stars, sorted by spectral type (earliest at the top, latest at the bottom) and normalised to their respective 7000 \AA\ fluxes. The spectra have been smoothed for clarity over a 1000-pixel range (18--24\AA). Strong molecular absorption bands of TiO and VO are visible in the cooler stars and have been labelled, along with strong telluric features. Note the strong flux from M54\,x01 below $\sim$7000 \AA\ when compared to the depth of its lines and molecular bands.}
\label{SpectraFig}
\end{figure*}

The data were taken using UVES (Ultra-Violet Echelle Spectrograph) on the European Southern Observatory Very Large Telescope (ESO VLT) at Cerro Paranal in Chile during 22--24 July 2003, with cross-disperser \#4. The spectra are in the wavelength range of 6100--9900 \AA, with a gap of $\sim$80 \AA\ around 8000 \AA\ due to the split between the two CCDs. Our observations were taken at a time of superb seeing ($\sim$0.3'' during the first 1.5 nights, increasing to 0.6'' for the remainder), allowing the use of the 0.3'' slit. Thus, the spectra are of very high resolution ($R = 110,000$). Data were reduced using the UVES package for {\sc midas} and are shown in Fig.~\ref{SpectraFig}.

Targets were selected from bright, red objects in the vicinity of galactic globular clusters and split into two categories based on the {\sc isocam} photometry of OFFR02: those with infra-red (IR) excess, $(K-[12]_0) \ge 0.65$, referred to in this paper by the `x' designator; and those without, $(K-[12]_0) < 0.65$, referred to by the `o' designator. The latter were selected from the 2MASS catalogue. A per-cluster summary of these objects can be found in Table \ref{ClusterTable}, and a list of individual objects in Table \ref{ObjectTable}. Radial velocities taken from the spectra show all the stars to be likely cluster members, with the exception of the star observed in M54. The target selection is biased towards stars near the tip of the RGB and AGB, i.e.\ the most luminous red giant stars in the clusters, although we show that they do comprise an order of magnitude range in bolometric luminosity. Variability could have resulted in a bias towards stars in the bright maximum of their pulsation cycle, but the IR amplitudes are only of order a tenth of a magnitude and well within the spread of the magnitudes. The inclusion of stars with known IR excess could in principle also bias the sample towards stars with higher mass-loss rates, but our analysis shows that stars without IR excess appear to experience similar mass-loss rates.

In two cases in NGC\,362 -- 2MASS\,J01031724$-$7050497 (B1) and 2MASS\,J01031474$-$7050589 (B2) -- the 2MASS data were not sufficient to determine which star was our target. In these cases, all possible counterparts (namely: o05a and o05b; and x02a, x02b and x02c, respectively) were also observed. Comparing the UVES acquisition and 2MASS images, we can see that the B1 is a blend comprising of two stars of similar magnitudes. We have assigned a flux based on Origlia et al.'s unpublished (J--K) data, split between the two stars on the basis of the comparative brightness in the acquisition images. The two stars have similar temperatures (see below), so we can expect them to have roughly similar (J--K) colours. B2 is a more complex blend of three bright stars (x02a, x02b and x02c) and two faint (unobserved) stars. The star x02c is more detached from the blend and has its own identifier (\#11) in Valenti et al.\ (2004, hereafter VFO04): we do not consider it to contribute significant flux to the 2MASS blend. We thus assume that the majority of the flux observed in the 2MASS object is from the stars x02a and x02b and split the flux as before. These stars were also found to have similar temperatures. It thus appears likely that the 2MASS/{\sc isocam} identifiers are better associated with o05b and x02a, respectively. We show later that both sources appear to have higher mass-loss rates than their companions.

Considerable literature already exists, both on the clusters and the individual stars we study. The cluster $\omega$ Centauri is well-noted for its chemically enriched sub-populations, with metallicities spanning over an order of magnitude between [Fe/H] $\sim$ --0.6 and [Fe/H] $<$ --1.6 (Norris et al.\ 1996; Pancino et al.\ 2000; Sollima et al.\ 2005; van Loon et al.\ 2007) and helium abundances of up to 38\% by fraction (Norris 2004). V42 is a well-known IR source (Origlia et al.\ 1995), studied recently spectroscopically by van Loon et al.\ (2007) and with the \emph{Spitzer Space Telescope} (Boyer et al.\ 2007, who find strong IR excess emission up to 24 $\mu$m). NGC\,6388, one of the most metal-rich clusters, may also have a helium-enriched population (Moehler \& Sweigart 2006). M\,15 is a very metal-poor cluster but signficantly the only cluster with a secure detection of ICM dust (Evans et al.\ 2003; Boyer et al.\ 2006) and gas (van Loon et al.\ 2006a). Boyer et al.\ find M\,15\,x01 is quiescent, but M\,15\,x02 to be mass-losing based on its IR colours. M\,54, a distant, but massive cluster with an unusually blue Horizontal Branch (Rosenberg et al.\ 2004) may be a component of the Sagittarius Dwarf Spheroidal galaxy (Ibata et al.\ 1994). NGC\,362 is a small globular in a direction near that of the Small Magellanic Cloud (SMC). Finally, 47\,Tuc, a more massive cluster near NGC\,362 and the SMC in the sky, has numerous variables which have recently been studied photometrically: Ramdani \& Jorissen discuss mass loss from V3 and V18, while Lebzelter \& Wood (2005) discovered variability in several of our sample (LW8, LW10, LW13 and LW19), and Ita et al.\ (2007) show V3 and LW13 to be on the giant branch tip, with V8 showing strong 11$\mu$m excess. Spectroscopic observations in the mid-IR have been carried out by Lebzelter et al.\ (2006) on V3, V8 and V18, and by van Loon et al.\ (2006b) on the above three stars and LW10, with possible silicate emission present in V18, which may have recently undergone a thermal pulse (Lebzelter et al.\ 2005).


\section{Stellar parameters}
\label{ParamsSection}

\subsection{Radial velocity}

The temperatures and radial velocities of our target stars were estimated using {\sc atlas9} model spectra (Kurucz 1993). A grid of model stellar spectra was created using a temperature range of 3500--6000 K, in 250 K steps; metallicity [Z/H] of $-2.0$ to $-0.5$, in steps of 0.5; and $\log(g)$ from 0 to 1.5, in steps of 0.5 dex. In addition to the normal atomic line lists, we included titanium oxide bands (Kurucz 1999) to better replicate spectra at lower temperatures.

A spectral type was assigned visually from standard reference spectra (Pickles 1998), and an {\sc atlas9} template spectrum was chosen from the above grid that best replicated the temperature expected for that spectral type, the host cluster metallicity, and an estimated gravity (based on Cioni et al.\ 2006a, 2006b). Radial velocities were found by cross-correlating the {\sc atlas9} spectrum with the stellar spectrum in regions rich in atomic or molecular lines and clear of sky lines (avoiding the H$\alpha$ and near-IR Ca II triplet lines). No significant difference between the velocities derived from the atomic- and molecular-line-rich regions was found. Spectra normalised in flux were also created using a running 1000-point (18--24 \AA) wide boxcar filter, chosen to minimise smoothing of molecular bands while maximising the removal of residual fringes that affect these red spectra.

For the hotter objects, this yielded radial velocities accurate to within $\sim$0.3 km s$^{-1}$, but for the cooler stars, where the strong molecular bands were not covered by the {\sc atlas9} models, the errors in radial velocity are much larger. In particular, those of 47\,Tuc\,x08, NGC\,6388\,o12 and $\omega$\,Cen\,x01 are over 10 km s$^{-1}$ (see Table \ref{ParamTable}). Comparison to other data (van Loon et al.\ 2007) shows that the velocity for $\omega$\,Cen\,x01 appears correct to within a few km s$^{-1}$. This suggests that the formal errors we have calculated may be too large.

The only object with a radial velocity not co-incident with its host cluster is that of M\,54\,x01, suggesting it is either a field star or in a compact binary with a relatively massive companion. The calculated luminosity for the distance of the cluster is not unreasonable for a red giant, and the spectrum appears to be that of a low gravity star. This suggests it may be in a binary system. However, attempts to cross-correlate the spectrum with the {\sc atlas9} models to find a binary component with a different radial velocity were unsuccessful and its velocity suggests that it is most likely a field star, possibly in the Galactic Bulge (M\,54 is at $l = 5.6^{\circ}, b = -14.1^{\circ}$).


\subsection{Stellar temperature}

\begin{figure}[!t]
\resizebox{\hsize}{!}{\includegraphics[angle=270]{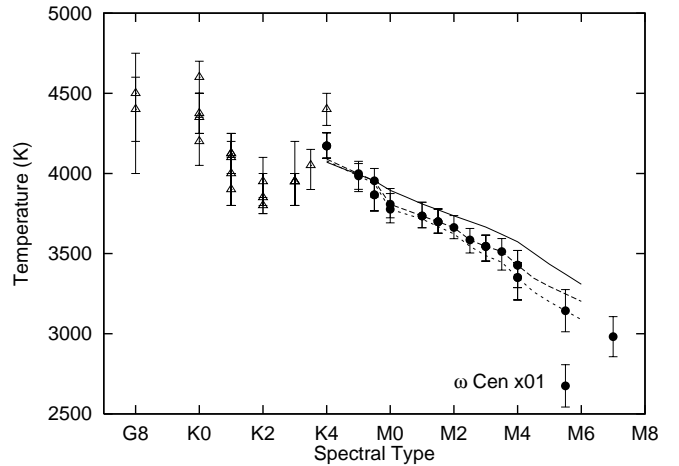}}
\caption[]{Comparison of temperature determinations between two methods: open triangles represent temperatures derived from Kurucz's {\sc atlas} 9 model fitting, filled circles represent `corrected' temperatures (see text) derived from Fluks et al.\ (1994) and Levesque et al.\ (2006). The lines show the corrections applied. The solid line represents Fluks et al.'s data for solar metallicity, the dashed line represents the correction to [Fe/H] $= -0.6$ (NGC\,6388) and the dotted line to [Fe/H] $= -0.76$ (47\,Tuc). The data for NGC\,362 at [Fe/H] $= -1.16$ have not been shown for clarity, as the correction is negligible in the observed range (bluewards of K5.5). The M5.5 outlier is $\omega$\,Cen\,x01, the fitting method for which is described in the text.}
\label{TempsFKFig}
\end{figure}

\begin{figure}[!btp]
\resizebox{\hsize}{!}{\includegraphics[angle=270]{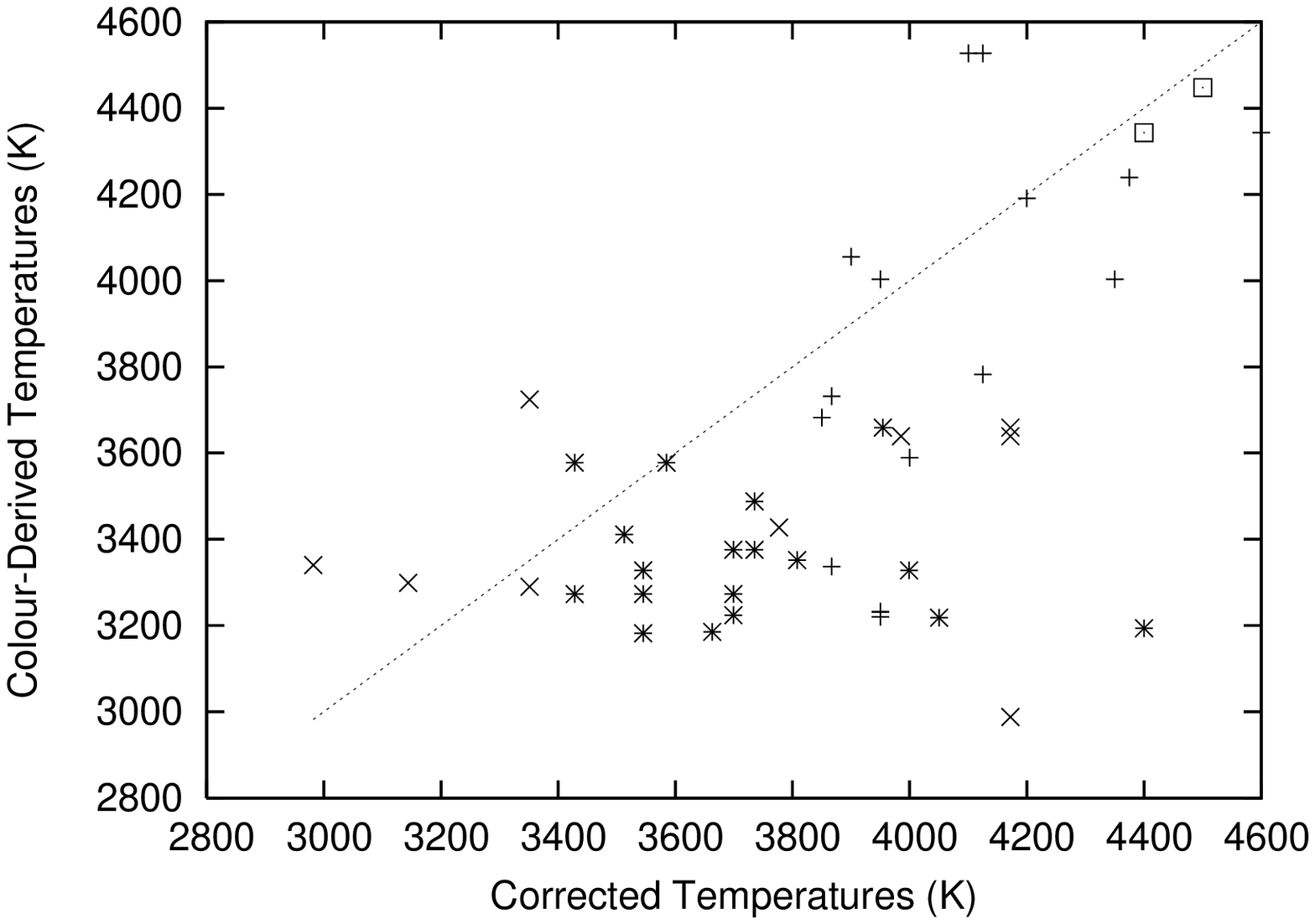}}
\caption[]{`Corrected' temperatures from spectral typing or {\sc atlas9} model fitting versus temperatures determined from (J--K) colour using Houdashelt et al.\ (2000). The two methods appear poorly correlated, as discussed in the text. Key: NGC\,362: plus-signs, 47\,Tuc: crosses, NGC\,6388:  asterisks, M\,15: open squares, $\omega$\,Cen: filled square.}
\label{TempTempsFig}
\end{figure}
Stellar temperatures were calculated for the hotter objects via chi-squared minimisation using the {\sc atlas9} models, again using the host cluster metallicity and estimated gravity. This proved unsuitable for the cooler objects due to limitations of the {\sc atlas9} models. For these objects (which typically have spectral class K4 or cooler), temperature was estimated from their spectral type using the temperature-spectral type correlation from Fluks et al.\ (1994). These correspond to the `uncorrected temperatures' in Table \ref{ParamTable}.

The Fluks et al.\ relations are valid for solar metallicity stars. As metallicity affects the value of effective temperature determined for a given spectral type, it was necessary to correct the cool objects' temperatures for the substantially sub-solar metallicities of the clusters. To do this, we used a similar relation for LMC, SMC and Galactic red supergiants from Levesque et al.\ (2006) and extrapolated the metallicity dependence at each spectral sub-type out to our clusters' metallicities using a second-order polynomial interpolation. We assumed [Z/H]$_{\mbox{\sc lmc}} = -0.3$ and [Z/H]$_{\mbox{\sc smc}} = -0.6$ (Russell \& Dopita 1990), and [Z/H]$_{\mbox{\sc mwg}} = 0$, and used a linear interpolation between sub-types where necessary. Overall, this has the effect of lowering the temperature of the stars by around 100--200 K.

This approach appears to be satisfactory in all cases except the M-giant in $\omega$\,Cen, for which this method would yield a temperature of 2000 K. Here, a linear extrapolation using the Milky Way and SMC data at spectral type M2 (the last datapoint in the Levesque et al.\ data) was used instead of a parabola. Due to the uncertainty of the extrapolation and possible errors introduced by this correction process, we suggest that the temperature we find is still rather low. Were this star part of the metal-rich sub-population, its temperature would be significantly higher (up to 3250 K, at [Z/H] $= -0.6$). We have, however, no evidence to support this theory, and van Loon et al.\ (2007) find that the M-type stars in $\omega$ Cen may have metallicities similar to the cluster average.

As the Levesque et al.\ data are for supergiants and not for the less massive and less luminous giants, we have to further correct for the difference. We perform a crude correction, using the offset between the temperatures of Levesque et al.'s supergiant galactic sample and Fluks et al.'s red giant (galactic) data. For each particular subtype, we subtract the difference from our result; again we interpolate where necessary. This has an effect of lowering the effective temperatures of types M0--M3 by $\sim$100 K, but little difference outside this range. These values correspond to the `corrected temperatures' in Table \ref{ParamTable}. They are plotted alongside the {\sc atlas9} model fits in Fig.\ \ref{TempsFKFig}.

A comparison between the two methods for deriving temperature shows that the {\sc atlas9} models are still consistently around 200 K cooler than the corrected temperatures from spectral typing (Fig.\ \ref{TempsFKFig}), suggesting our corrections have not been as large as necessary. This may be due to the uncertainty in extrapolating the empirical corrections to low metallicities. One would presume this offset still exists at cooler temperatures where the {\sc atlas9} models have not made a fit, however as this is uncertain we have not made any further corrections for this in our results. This suggests that metal-poor spectra may correspond to even lower temperatures (around the lower error values) than calculated here.

Comparisons were made with the colour-temperature tables in Houdashelt et al.\ (2000), using (J--K) colour to estimate a temperature. The temperatures derived using this method are plotted against the temperatures found using the above methods (after corrections) in Fig.\ \ref{TempTempsFig}. The Houdashelt et al.\ relations are calibrated only for temperatures $\geq$ 4000 K, thus it is not surprising that we observe a poor correlation between the two methods.


\subsection{Physical Parameters}

Physical parameters of the stars have also been calculated (Table \ref{ParamTable}). Luminosities were found from K-band magnitudes and using bolometric corrections made from Montegriffo et al.\ (1998) using a linear interpolation for metallicity. Radii were then computed using the `corrected temperatures' described above and, using these, escape velocities were also computed. A mass of 0.8 M$_{\odot}$ was assumed for all stars, though some stars near the AGB tip may have lost up to $\sim$30\% of this, which would lower their escape velocities by up to $\sim$17\%.


\subsection{Analysis}

\begin{table*}[!htp]
\begin{center}
\caption{Stellar parameters derived from spectral typing and comparisons to model atmospheres.}
\label{ParamTable}
\begin{tabular}{l@{\ }l@{\ }cr@{\,}lr@{\,}l@{\ }c@{\ }r@{\,}lr@{\,}lr@{\,}lr@{\,}l}
    \hline \hline
ID  & \multicolumn{1}{c}{Spectral}  & *        &\multicolumn{2}{c}{Uncorrected}    &\multicolumn{2}{c}{Corrected}   &Colour-Derived	&\multicolumn{2}{c}{Radial}       	&\multicolumn{2}{c}{Luminosity}	&\multicolumn{2}{c}{Radius}		&\multicolumn{2}{c}{Escape}\\
    & \multicolumn{1}{c}{Type}      &       &\multicolumn{2}{c}{Temperature}    &\multicolumn{2}{c}{Temperature} &Temperature	&\multicolumn{2}{c}{Velocity}     	&	&	&	&		&\multicolumn{2}{c}{Velocity}\\
    &      			    &        &\multicolumn{2}{c}{(K)}            &\multicolumn{2}{c}{(K)}         &(K)	&\multicolumn{2}{c}{(km s$^{-1}$)}	&\multicolumn{2}{c}{(L$_{\odot}$)}	&\multicolumn{2}{c}{(R$_{\odot}$)}	&\multicolumn{2}{c}{(km s$^{-1}$)}\\
    \hline 

\multicolumn{16}{c}{NGC\,362} \\
o01 & \ \ K5.5 & F & 3954 & $^{+  77 }_{- 101 }$ & 3867 & $^{+  77 }_{- 101 }$ & 3337 &     2.5 & $\pm$  4.3 &  2415 & $\pm$ 102 & 109 & $\pm$  5 & 53 & $\pm$  6 \\[-0.0pt]
o02 & \ \ K0   & K & 4200 & $^{+ 250 }_{- 250 }$ & 4200 & $^{+ 250 }_{- 250 }$ & 4191 &     9.3 & $\pm$  1.0 &   988 & $\pm$  86 &  59 & $\pm$  6 & 72 & $\pm$ 10 \\[-0.0pt]
o03 & \ \ K1   & K & 3900 & $^{+ 250 }_{- 250 }$ & 3900 & $^{+ 250 }_{- 250 }$ & 4055 &  --16.7 & $\pm$  1.0 &  2327 & $\pm$ 226 & 105 & $\pm$ 12 & 54 & $\pm$  8 \\[-0.0pt]
o04 & \ \ K1   & K & 4125 & $^{+ 250 }_{- 250 }$ & 4125 & $^{+ 250 }_{- 250 }$ & 3782 &   --4.2 & $\pm$  0.3 &  1468 & $\pm$ 131 &  75 & $\pm$  8 & 64 & $\pm$  9 \\[-0.0pt]
o05a& \ \ K1   & K & 4100 & $^{+ 250 }_{- 250 }$ & 4100 & $^{+ 250 }_{- 250 }$ & 4527 &   --6.8 & $\pm$  0.3 &  2103 & $\pm$ 189 &  91 & $\pm$ 10 & 58 & $\pm$  9 \\[-0.0pt]
o05b& \ \ K1   & K & 4125 & $^{+ 250 }_{- 250 }$ & 4125 & $^{+ 250 }_{- 250 }$ & 4527 &   --5.8 & $\pm$  0.3 &  2392 & $\pm$ 213 &  95 & $\pm$ 10 & 57 & $\pm$  8 \\[-0.0pt]
o06 & \ \ K5.5 & F & 3954 & $^{+  77 }_{- 101 }$ & 3867 & $^{+  77 }_{- 101 }$ & 3732 &     0.7 & $\pm$  2.7 &  2505 & $\pm$ 105 & 111 & $\pm$  6 & 52 & $\pm$  6 \\[-0.0pt]
o07 & \ \ K3   & K & 3950 & $^{+ 250 }_{- 250 }$ & 3950 & $^{+ 250 }_{- 250 }$ & 4003 &   --8.6 & $\pm$  0.3 &  2327 & $\pm$ 222 & 103 & $\pm$ 12 & 54 & $\pm$  8 \\[-0.0pt]
o08 & \ \ K2   & K & 3850 & $^{+ 250 }_{- 250 }$ & 3850 & $^{+ 250 }_{- 250 }$ & 3682 &     7.3 & $\pm$  0.3 &  2182 & $\pm$ 215 & 105 & $\pm$ 12 & 54 & $\pm$  8 \\[-0.0pt]
o09 & \ \ K0   & K & 4350 & $^{+ 250 }_{- 250 }$ & 4350 & $^{+ 250 }_{- 250 }$ & 4003 &    10.3 & $\pm$  0.3 &   665 & $\pm$  54 &  45 & $\pm$  5 & 82 & $\pm$ 12 \\[-0.0pt]
o10 & \ \ K0   & K & 4600 & $^{+ 250 }_{- 250 }$ & 4600 & $^{+ 250 }_{- 250 }$ & 4343 &     1.7 & $\pm$  0.4 &   659 & $\pm$  48 &  40 & $\pm$  4 & 87 & $\pm$ 12 \\[-0.0pt]
x01 & \ \ K1   & K & 4000 & $^{+ 250 }_{- 250 }$ & 4000 & $^{+ 250 }_{- 250 }$ & 3589 &   --1.6 & $\pm$  0.3 &  1866 & $\pm$ 175 &  90 & $\pm$ 10 & 58 & $\pm$  9 \\[-0.0pt]
x02a& \ \ K3   & K & 3950 & $^{+ 250 }_{- 250 }$ & 3950 & $^{+ 250 }_{- 250 }$ & 3232 &   --1.4 & $\pm$  0.3 &  2599 & $\pm$ 248 & 109 & $\pm$ 12 & 53 & $\pm$  8 \\[-0.0pt]
x02b& \ \ K3   & K & 3950 & $^{+ 250 }_{- 250 }$ & 3950 & $^{+ 250 }_{- 250 }$ & 3232 &     6.2 & $\pm$  0.4 &  2824 & $\pm$ 269 & 113 & $\pm$ 13 & 52 & $\pm$  8 \\[-0.0pt]
x02c& \ \ K0   & K & 4375 & $^{+ 250 }_{- 250 }$ & 4375 & $^{+ 250 }_{- 250 }$ & 4239 &     0.9 & $\pm$  0.3 &  2027 & $\pm$ 164 &  78 & $\pm$  8 & 62 & $\pm$  9 \\[-0.0pt]
x03 & \ \ K2   & K & 3950 & $^{+ 250 }_{- 250 }$ & 3950 & $^{+ 250 }_{- 250 }$ & 3220 &   --4.1 & $\pm$  0.3 &  2182 & $\pm$ 208 &  99 & $\pm$ 11 & 55 & $\pm$  8 \\
    \hline             
\multicolumn{16}{c}{47\,Tuc} \\             
x01 & \ \ M4   & F & 3574 & $^{+  92 }_{- 140 }$ & 3351 & $^{+  92 }_{- 140 }$ & 3724 &   --7.4 & $\pm$  6.4 &  6172 & $\pm$ 427 & 232 & $\pm$ 19 & 36 & $\pm$  5 \\[-0.0pt]
x02 & \ \ K4   & F & 4147 & $^{+  82 }_{-  77 }$ & 4172 & $^{+  82 }_{-  77 }$ & 3659 &     7.2 & $\pm$  2.4 &  1375 & $\pm$  51 &  71 & $\pm$  3 & 66 & $\pm$  7 \\[-0.0pt]
x03 & \ \ K5   & F & 3993 & $^{+  77 }_{-  98 }$ & 3985 & $^{+  77 }_{-  98 }$ & 3639 &  --10.5 & $\pm$  0.6 &  2325 & $\pm$ 103 & 101 & $\pm$  5 & 55 & $\pm$  6 \\[-0.0pt]
x04 & \ \ K4   & F & 4147 & $^{+  82 }_{-  77 }$ & 4172 & $^{+  82 }_{-  77 }$ & 3639 &   --5.6 & $\pm$  3.1 &  1427 & $\pm$  53 &  72 & $\pm$  3 & 65 & $\pm$  7 \\[-0.0pt]
x05 & \ \ M0   & F & 3895 & $^{+  98 }_{-  85 }$ & 3777 & $^{+  98 }_{-  85 }$ & 3427 &    10.4 & $\pm$  3.0 &  3180 & $\pm$ 135 & 131 & $\pm$  6 & 48 & $\pm$  5 \\[-0.0pt]
x06 & \ \ M5.5 & F & 3371 & $^{+ 132 }_{- 132 }$ & 3144 & $^{+ 132 }_{- 132 }$ & 3299 &   --2.7 & $\pm$  7.3 &  6463 & $\pm$ 448 & 270 & $\pm$ 22 & 34 & $\pm$  4 \\[-0.0pt]
x07 & \ \ M4   & F & 3574 & $^{+  92 }_{- 140 }$ & 3351 & $^{+  92 }_{- 140 }$ & 3290 &     4.5 & $\pm$  6.4 &  7352 & $\pm$ 509 & 254 & $\pm$ 21 & 35 & $\pm$  4 \\[-0.0pt]
x08 & \ \ M7   & F & 3188 & $^{+ 125 }_{- 125 }$ & 2982 & $^{+ 125 }_{- 125 }$ & 3340 &   --6.5 & $\pm$ 11.7 & 10338 & $\pm$ 711 & 380 & $\pm$ 31 & 28 & $\pm$  4 \\[-0.0pt]
x09 & \ \ K4   & F & 4147 & $^{+  82 }_{-  77 }$ & 4172 & $^{+  82 }_{-  77 }$ & 2988 &    10.8 & $\pm$  2.6 &  4270 & $\pm$ 158 & 125 & $\pm$  5 & 49 & $\pm$  5 \\
    \hline             
\multicolumn{16}{c}{NGC\,6388} \\             
o01 & \ \ M4   & F & 3574 & $^{+  92 }_{- 140 }$ & 3428 & $^{+  92 }_{- 140 }$ & 3578 &     4.4 & $\pm$  4.1 &  1820 & $\pm$ 117 & 121 & $\pm$  9 & 50 & $\pm$  6 \\[-0.0pt]
o02 & \ \ M3   & F & 3666 & $^{+  70 }_{-  92 }$ & 3545 & $^{+  70 }_{-  92 }$ & 3328 &  --16.0 & $\pm$  4.0 &  4055 & $\pm$ 171 & 168 & $\pm$  8 & 43 & $\pm$  5 \\[-0.0pt]
o03 & \ \ K3   & K & 4050 & $^{+ 250 }_{- 250 }$ & 4050 & $^{+ 250 }_{- 250 }$ & 3218 &    23.8 & $\pm$  0.3 &  1331 & $\pm$ 122 &  74 & $\pm$  8 & 64 & $\pm$ 10 \\[-0.0pt]
o04 & \ \ M0   & F & 3895 & $^{+  98 }_{-  85 }$ & 3808 & $^{+  98 }_{-  85 }$ & 3352 &    31.7 & $\pm$  3.0 &  2168 & $\pm$  79 & 107 & $\pm$  5 & 53 & $\pm$  6 \\[-0.0pt]
o05 & \ \ K5   & F & 3993 & $^{+  77 }_{-  98 }$ & 3999 & $^{+  77 }_{-  98 }$ & 3328 &     3.4 & $\pm$  3.1 &  1380 & $\pm$  54 &  77 & $\pm$  3 & 63 & $\pm$  7 \\[-0.0pt]
o06 & \ \ M1   & F & 3810 & $^{+  85 }_{-  74 }$ & 3735 & $^{+  85 }_{-  74 }$ & 3488 &     3.1 & $\pm$  2.5 &  2312 & $\pm$  77 & 115 & $\pm$  4 & 52 & $\pm$  6 \\[-0.0pt]
o07 & \ \ K4   & K & 4400 & $^{+ 250 }_{- 250 }$ & 4400 & $^{+ 250 }_{- 250 }$ & 3194 &     7.2 & $\pm$  0.8 &  1445 & $\pm$ 115 &  65 & $\pm$  6 & 68 & $\pm$ 10 \\[-0.0pt]
o08 & \ \ M2   & F & 3736 & $^{+  74 }_{-  70 }$ & 3663 & $^{+  74 }_{-  70 }$ & 3185 &  --19.5 & $\pm$  3.8 &  2032 & $\pm$  66 & 112 & $\pm$  4 & 52 & $\pm$  6 \\[-0.0pt]
o09 & \ \ M1.5 & F & 3773 & $^{+  79 }_{-  72 }$ & 3699 & $^{+  79 }_{-  72 }$ & 3376 &     5.9 & $\pm$  3.0 &  1077 & $\pm$  35 &  80 & $\pm$  3 & 62 & $\pm$  7 \\[-0.0pt]
o10 & \ \ K5.5 & F & 3954 & $^{+  77 }_{- 101 }$ & 3954 & $^{+  77 }_{- 101 }$ & 3659 &  --16.4 & $\pm$  2.6 &  1331 & $\pm$  54 &  78 & $\pm$  4 & 63 & $\pm$  7 \\[-0.0pt]
o11 & \ \ M3.5 & F & 3620 & $^{+  81 }_{- 116 }$ & 3513 & $^{+  81 }_{- 116 }$ & 3411 &     0.1 & $\pm$  8.4 &  1406 & $\pm$  74 & 101 & $\pm$  6 & 55 & $\pm$  6 \\[-0.0pt]
o12 & \ \ M4   & F & 3574 & $^{+  92 }_{- 140 }$ & 3428 & $^{+  92 }_{- 140 }$ & 3273 &  --12.5 & $\pm$ 11.3 &  1722 & $\pm$ 111 & 117 & $\pm$  9 & 51 & $\pm$  6 \\
x01 & \ \ M3.5 & F & 3701 & $^{+  72 }_{-  81 }$ & 3585 & $^{+  72 }_{-  81 }$ & 3578 &   --0.1 & $\pm$  4.4 &  1644 & $\pm$  61 & 105 & $\pm$  5 & 54 & $\pm$  6 \\[-0.0pt]
x02 & \ \ M1   & F & 3810 & $^{+  85 }_{-  74 }$ & 3735 & $^{+  85 }_{-  74 }$ & 3376 &   --5.7 & $\pm$  2.4 &  1923 & $\pm$  64 & 104 & $\pm$  4 & 54 & $\pm$  6 \\[-0.0pt]
x03 & \ \ M1.5 & F & 3773 & $^{+  79 }_{-  72 }$ & 3699 & $^{+  79 }_{-  72 }$ & 3273 &  --19.6 & $\pm$  3.4 &  3076 & $\pm$ 101 & 135 & $\pm$  5 & 48 & $\pm$  5 \\[-0.0pt]
x04 & \ \ M3   & F & 3666 & $^{+  70 }_{-  92 }$ & 3545 & $^{+  70 }_{-  92 }$ & 3273 &    26.4 & $\pm$  4.0 &  1514 & $\pm$  64 & 103 & $\pm$  5 & 54 & $\pm$  6 \\[-0.0pt]
x05 & \ \ M1.5 & F & 3773 & $^{+  79 }_{-  72 }$ & 3699 & $^{+  79 }_{-  72 }$ & 3224 &     7.9 & $\pm$  2.4 &  2421 & $\pm$  79 & 119 & $\pm$  5 & 51 & $\pm$  5 \\[-0.0pt]
x06 & \ \ M3   & F & 3666 & $^{+  70 }_{-  92 }$ & 3545 & $^{+  70 }_{-  92 }$ & 3182 &     9.2 & $\pm$  3.6 &  2831 & $\pm$ 119 & 141 & $\pm$  7 & 47 & $\pm$  5 \\[-0.0pt]
    \hline             
\multicolumn{16}{c}{M\,15} \\             
x01 & \ \ G8   & K & 4500 & $^{+ 250 }_{- 250 }$ & 4500 & $^{+ 250 }_{- 250 }$ & 4448 &  --14.7 & $\pm$  7.7 &   489 & $\pm$  37 &  36 & $\pm$  3 & 92 & $\pm$ 13 \\[-0.0pt]
x02 & \ \ G8   & K & 4400 & $^{+ 250 }_{- 250 }$ & 4400 & $^{+ 250 }_{- 250 }$ & 4343 &    10.3 & $\pm$  5.1 &   891 & $\pm$  71 &  51 & $\pm$  5 & 77 & $\pm$ 11 \\
    \hline             
\multicolumn{16}{c}{M\,54} \\             
x01 & \ \ K2   & K & 3800 & $^{+ 250 }_{- 250 }$ & 3800 & $^{+ 250 }_{- 250 }$ &      & --182.9 & $\pm$  0.2 &       &           &     &          &    &          \\
    \hline             
\multicolumn{16}{c}{$\omega$\,Cen} \\             
x01 & \ \ M5.5 & F & 3371 & $^{+ 132 }_{- 132 }$ & 2675 & $^{+ 132 }_{- 132 }$ & 3707 &    44.3 & $\pm$ 26.2 &  2111 & $\pm$ 161 & 213 & $\pm$ 19 & 38 & $\pm$  5 \\
    \hline
\multicolumn{16}{p{0.83\textwidth}}{\small Notes: Column 3: F -- from spectral typing, using Fluks et al.\ (1994); K -- derived from Kurucz's {\sc atlas9} models (see Section 3). The colour-derived temperature is from (J--K) colours, from Houdashelt et al.\ (2000) (see text). Radial velocity is relative to the mean motion for the cluster from Harris (1996), listed in Table \ref{ClusterTable}. See text for details on M54\,x01 and $\omega$\,Cen\,x01. Formal errors have been assigned on the basis of a 0.1 dex uncertainty in metallicity, a 100 pc uncertainty in distance and a 0.1 M$_{\odot}$ error in mass. \normalsize} 
\\
    \hline
\\
\end{tabular}
\end{center}
\end{table*}

\begin{figure}[!btp]
\resizebox{\hsize}{!}{\includegraphics[angle=270]{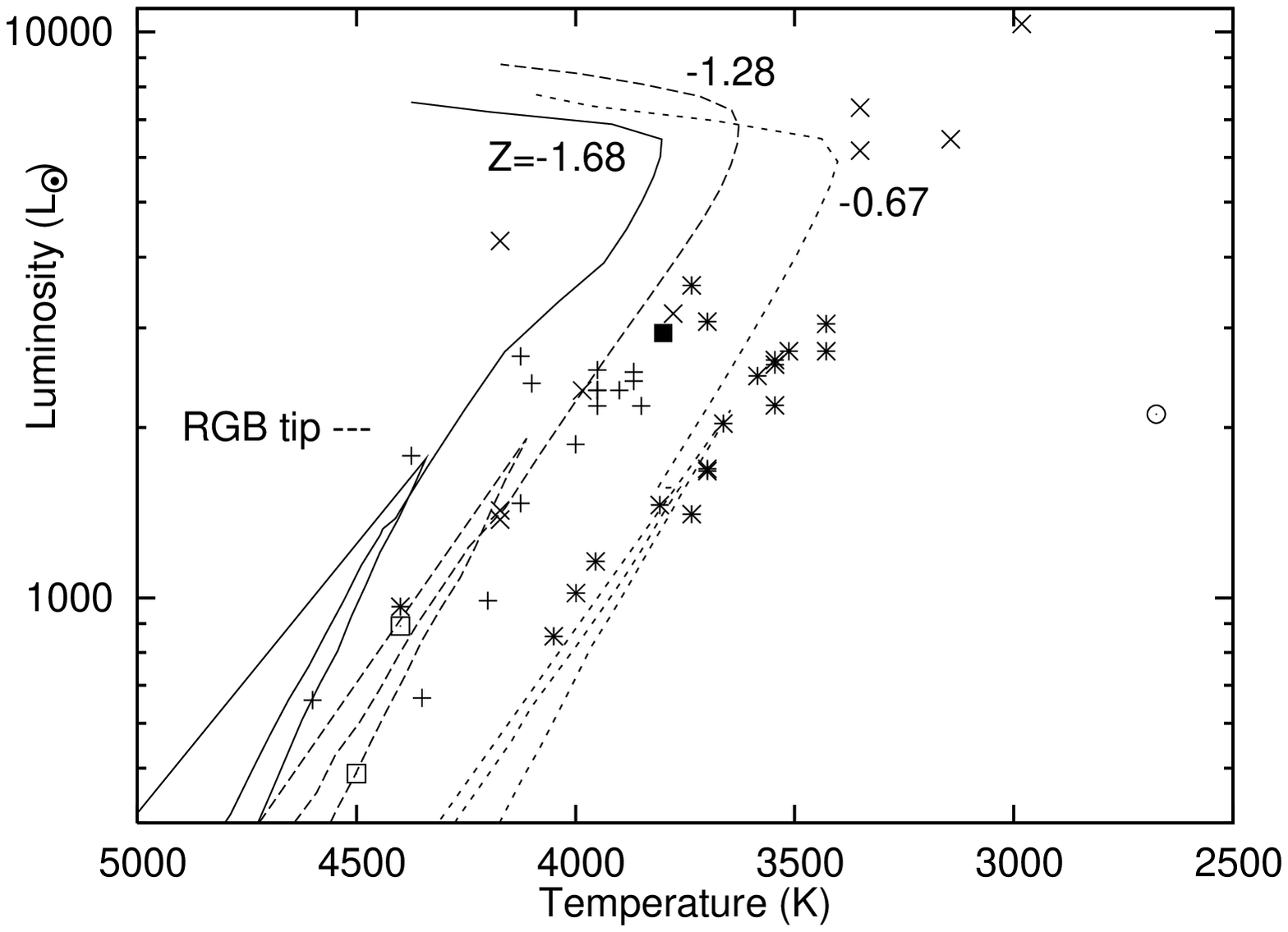}}
\resizebox{\hsize}{!}{\includegraphics[angle=270]{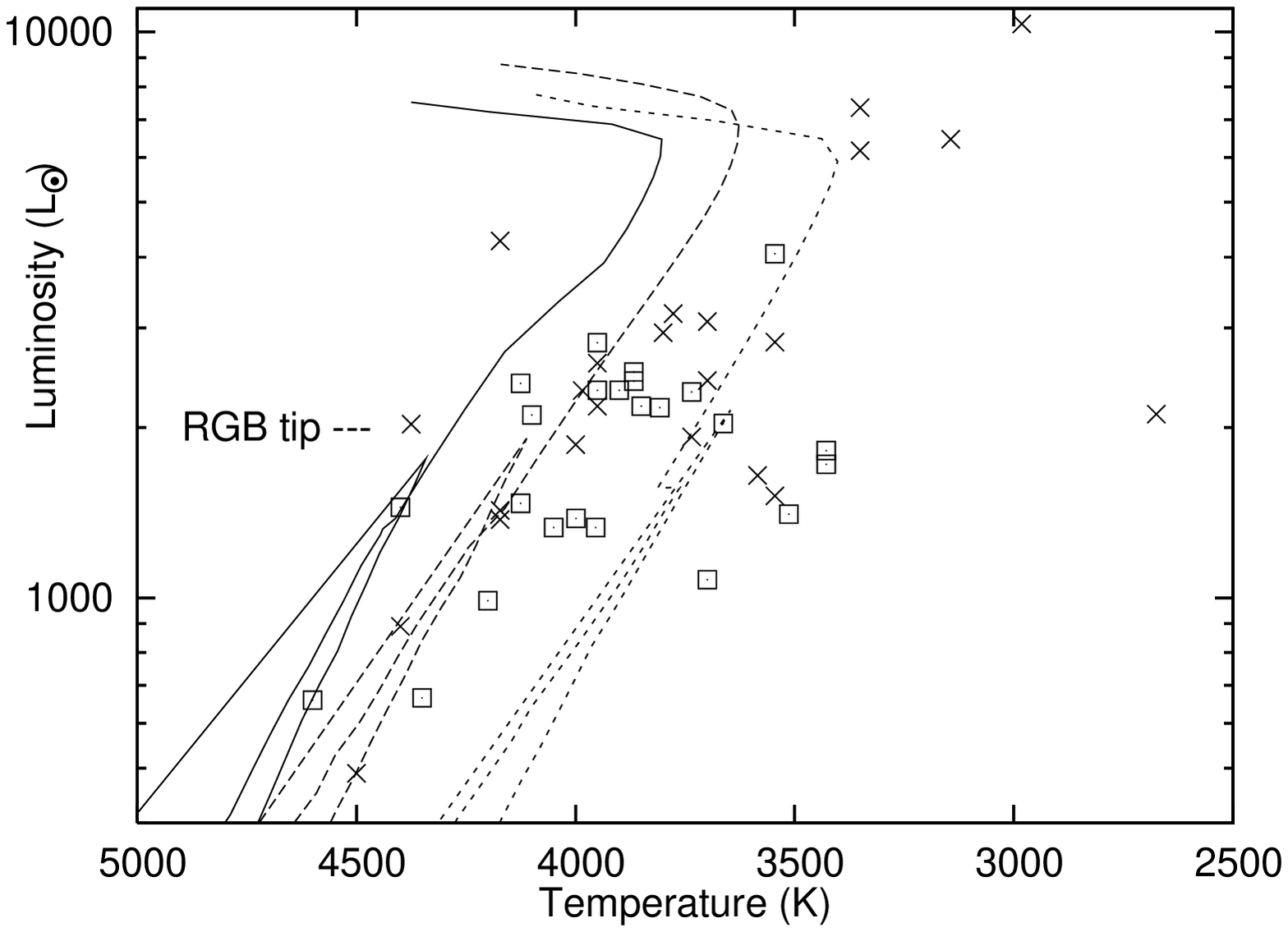}}
\caption[]{Hertzsprung-Russell diagrams of our targets. Top panel: stars identified by cluster: NGC\,362 -- plus signs, 47\,Tuc -- crosses, NGC\,6388 -- asterisks, M\,15 -- boxes, M\,54 -- filled box, $\omega$\,Cen -- circle. Bottom panel: IR excessive stars (crosses) and non-excessive stars (boxes). The temperature for our $\omega$\,Cen target has probably been underestimated. Also shown are isochrones for [Z/H] = --0.67, --1.28 and --1.68 (labelled on top panel) from the models of Cioni et al.\ (2006a, 2006b) at an age of 12.6 Gyr.}
\label{UVESHRD}
\end{figure}

We display a Hertzsprung-Russell diagram in Fig.\ \ref{UVESHRD}, including isochrones at 12.6 Gyr from Cioni et al.\ (2006a, 2006b). There is significant scatter in the plots, likely due to the uncertainty in temperature of the stars, though the stars of metal-rich and metal-poor clusters clearly match the metal-rich and metal-poor isochrones, respectively. The position of the RGB tip, at around 2000 L$_{\odot}$, does not change appreciably with metallicity nor within the likely range of ages. Assuming, therefore, that the distances to the clusters are correct, we expect that the stars above the RGB tip are on the AGB (below the tip it is difficult to tell which branch they belong to) suggesting more than half of the sample is on the AGB. Note that, due to their intrinsic variability, the stars will naturally display a range in both luminosity and temperature: a 0.5 (bolometric) magnitude variability would change the luminosity by a factor of 1.58, and the temperature by about 130 K (assuming the net flux at the surface is roughly constant).

\begin{figure}[!btp]
\resizebox{\hsize}{!}{\includegraphics[angle=270]{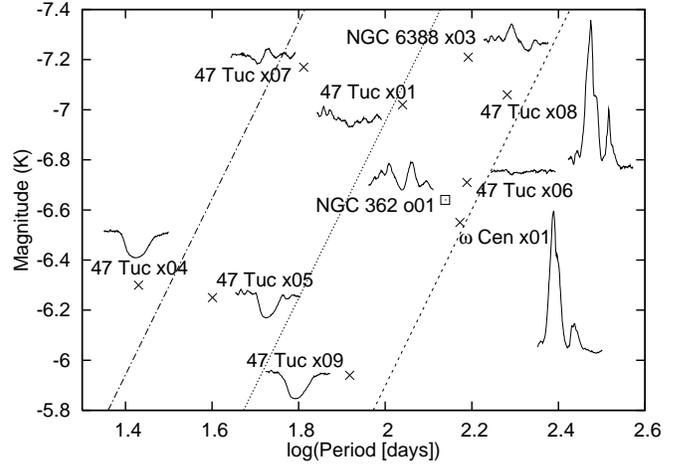}}
\caption[]{Period-magnitude diagram for stars with variability data. The lines show the fundamental mode (longest period), and first and second overtone (harmonic) periods, using the relations for the LMC given by Ita et al.\ (2004). The corresponding H$\alpha$ profiles are plotted near the positions of the stars. Clearly, H$\alpha$ emission is strongest in the fundamental mode and is therefore most likely linked to shocks in the pulsating atmosphere.}
\label{UVESPL}
\end{figure}

A period-magnitude diagram is shown in Fig.\ \ref{UVESPL}, using periodicity data from the Updated Third Catalogue of Variable Stars in Globular Clusters (Clement 1997) and from Matsunaga et al.\ (2007), Ita et al.\ (2007) and Matsunaga (private communication). Frequency modes from Ita et al.\ (2004) have also been shown. These relations are for the LMC, so we expect our more metal-poor sample to be slightly brighter for a given period (Wood 1990). It is notable that very few stars in our sample have known variability data, and we discuss this in Sect.\ 6.1.

\section{H$\alpha$ and Ca II profiles}

\begin{figure*}[!pbt]
\resizebox{\hsize}{!}{\includegraphics{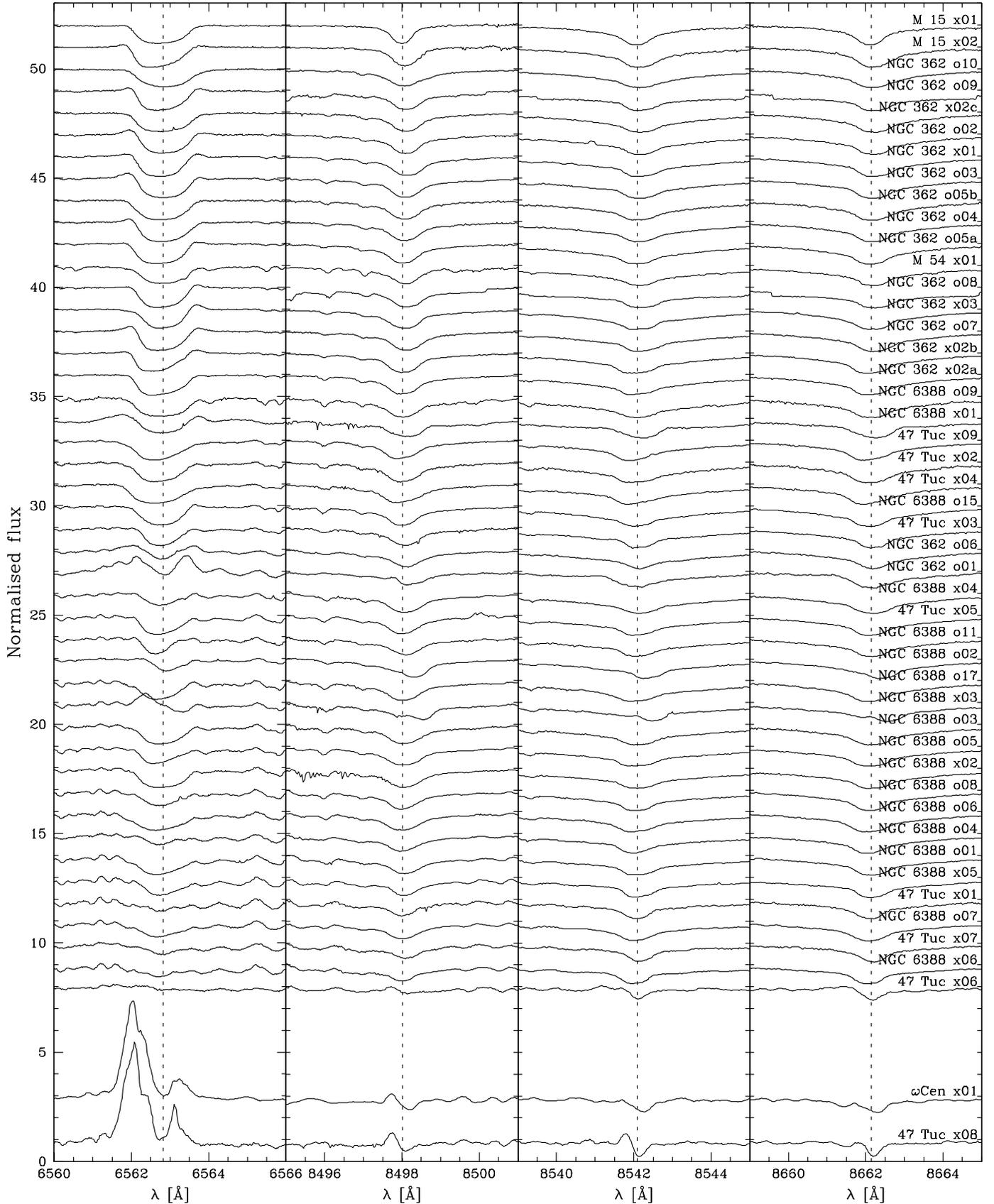}}
\caption[]{H$\alpha$ and near-IR calcium triplet line profiles of the target stars, sorted by spectral type as in Fig.~\ref{SpectraFig} and normalised to their continuum fluxes at the wavelengths of the respective lines. The spectra have also been shifted to laboratory wavelengths. Note the variety of asymmetric emission and absorption in the lines, especially in H$\alpha$. We consider the strong H$\alpha$ emission from the two latest-type objects to be shock emission from pulsation. The similarity of the profiles is remarkable, despite very different metallicities.}
\label{HalphaFig}
\end{figure*}

In an effort to find mass-losing stars, we investigated the H$\alpha$ (6562.8 \AA) and calcium IR triplet lines (8498, 8542 and 8662 \AA) for evidence of asymmetry and velocity shift. These lines are shown in Fig.\ \ref{HalphaFig}. To accomplish this, we calculated line bisectors by interpolating the line profiles to uniform steps in flux. At each given flux, the two points at either side of the line centre (as determined by the line's rest wavelength with respect to the calculated radial velocity) were averaged.

Examination of the calcium triplet line bisectors shows most to be fairly symmetrical. Some show slight blue- or red-shifted cores of the order 1--3 km s$^{-1}$, but these are not usually visible in all three lines and can probably be attributed to statistical noise, rather than as certain indicators of mass loss or pulsation (the exceptions being 47\,Tuc\,x04 and x09, which show 4--5 km s$^{-1}$ blue-shifted cores in all three lines). We visually estimate that the uncertainty on the shifts is of the order of 0.5--2 km s $^{-1}$, depending on the signal-to-noise of the data and effects of intruding metal lines. This is similar to, or less than, the radial velocity uncertainties. The H$\alpha$ line shows similar features in some cases, but overall shows much more prominent emission and more significantly blue-shifted absorption at velocities of 2--12 km s$^{-1}$. The H$\alpha$ lines also tend to be very deep, with over 90\% absorption in the core in some cases.

In both elements, the lines were predominantly found to be `banana-shaped', with a reddened kink at approximately 30--50\% of the peak absorption depth. As such, the profiles are not well suited to the usual method of finding a `velocity shift'; thus, we calculated the radial velocities of these `kink' points and of the line cores (negating anything appearing to be statistical scatter due to Poisson noise) with respect to the stellar velocities described above. The results are included in Table \ref{ShiftsTable}.

Notable exceptions to the above features appear in generally more luminous and cooler objects. Strong emission is clearly present in 47\,Tuc\,x08 and $\omega$\,Cen\,x01 in H$\alpha$ and (to a lesser extent) in Ca II. Lesser versions of this can also be seen in NGC\,362\,o01 and to some extent in NGC\,6388\,x03 (Fig.\ 5). These four stars show a much larger K-band variability (hence much stronger pulsations) than the remainder of the objects for which we have variability data (see Table \ref{ObjectTable}). This emission is therefore almost certainly due to heating by shock waves caused by pulsations. NGC\,362\,o01 and $\omega$\,Cen\,x01 have also been shown to have variable Balmer line emission (Lloyd Evans 1983a, 1983b), further suggesting this is pulsation related. 47\,Tuc\,x09 also shows variable H$\alpha$ absorption (Lloyd Evans 1984).

Furthermore, it is interesting to note that the two most extreme objects show qualitatively very similar profiles in H$\alpha$ and to a lesser extent in the calcium triplet. This is particularly interesting given their clearly different metallicities: 47\,Tuc\,x08 is estimated to be at 3--8 times of the metallicity of $\omega$\,Cen\,x01, which is visibly manifest in the other, stronger spectral lines of the former in Fig.\ \ref{HalphaFig}. This has obvious implications for the variance (or otherwise) of mass-loss rate with metallicity.
 
Weaker emission is also apparent in warmer stars. This has been noted for many giant stars in the literature and two related explanations have been put forward to model this. Cohen (1976) assumes a cooling circumstellar envelope surrounding the star at around two stellar radii. Another theory, by Dupree et al.\ (1984), states that H$\alpha$ emission could arise from chromospheric emission from a similar, hot shell source. Cacciari et al.~(2004) have suggested that active chromospheres are the dominant producer of line emission in these stars. Mauas et al.\ (2006) then used these chromospheric lines to model RGB mass loss in good agreement with evolutionary theory and previous estimates (Dupree et al.\ 1990, 1994) -- notably they find the need for an outward velocity field in the chromosphere, which we also find (see discussion).

It is an obvious suggestion, backed up by more detailed modelling by Dupree et al.\ and Mauas et al., that blue-shifted absorption cores are evidence of outflow from the stars. The fact that we see red-shifted cores in only two cases (two of the pulsation-shocked objects) suggests that there is little or no inflow to the stars, or that this only occurs at very particular phases of the pulsation cycle. We assume that there is therefore a net mass loss from the sample. 


\begin{table*}[!Htbp]
\begin{center}
\caption{Radial velocities of the H$\alpha$ and Ca II lines as described in Sect.\ 4.}
\label{ShiftsTable}
\begin{tabular}{ll@{\qquad\qquad},,,,,,,,}
    \hline \hline
ID     & \multicolumn{1}{c}{H$\alpha$}  & \multicolumn{2}{c}{H$\alpha$ velocity}  &  \multicolumn{2}{c}{Ca II velocity}  & \multicolumn{2}{c}{Ca II velocity}  & \multicolumn{2}{c}{Ca II velocity}  \\
       & \multicolumn{1}{c}{profile}    & \multicolumn{2}{c}{6563 \AA} & \multicolumn{2}{c}{8498 \AA} & \multicolumn{2}{c}{8542 \AA} & \multicolumn{2}{c}{8662 \AA} \\
       & \multicolumn{1}{c}{appearance} & \multicolumn{1}{c}{Core} & \multicolumn{1}{c}{Kink} & \multicolumn{1}{c}{Core} & \multicolumn{1}{c}{Kink} & \multicolumn{1}{c}{Core} & \multicolumn{1}{c}{Kink} & \multicolumn{1}{c}{Core} & \multicolumn{1}{c}{Kink} \\
    \hline
\multicolumn{10}{c}{NGC\,362} \\
o01	&	MOL, BE, RE	&	3.3^c	&	-3.1^{ca}&	4.0	&	10.5	&	-0.9	&	10.9	&	-0.6	&	6.7\\
o02	&	PHOT, be	&	-1.9:	&	4.8^a	&	2.7^d	&	0.6^d	&	2.2^b	&	2.2^b	&	1.9^b	&	1.9^b	\\
o03	&	PHOT, be	&	0.7	&	5.0^a	&	1.7	&	3.5	&	1.3	&	4.0	&	1.3	&	3.9\\
o04	&	PHOT, be	&	-2.4:	&	6.5^a	&	0.5	&	1.7	&	0.7	&	2.7	&	0.1	&	2.1\\
o05a	&	PHOT, re?	&	-6.5	&	-0.4	&	1.5^b	&	1.5^b	&	0.9^b	&	0.9^b	&	0.0	&	2.0\\
o05b	&	PHOT, wide	&	-1.8:	&	0.9^a	&	2.0^b	&	2.0^b	&	0.0	&	2.0	&	0.5	&	2.9\\
o06	&	phot, BE, RE	&	-1.6:	&	2.5^a	&	2.8^b	&	2.8^b	&	1.0	&	2.2	&	1.3	&	4.2\\
o07	&	PHOT, be, re	&	-6.2	&	4.1^a	&	3.5^b	&	3.5^b	&	-0.8	&	4.8	&	3.3^b	&	3.3^b	\\
o08	&	PHOT, re	&	-7.9:	&	-0.1	&	1.3	&	1.8	&	0.8	&	2.0	&	1.5	&	2.5\\
o09	&	PHOT, be, re	&	-5.4:	&	2.0^a	&	-0.5	&	1.4	&	-1.9	&	2.6	&	1.0	&	1.5\\
o10	&	PHOT		&	2.9	&	9.3^a	&	2.1	&	5.1	&	2.9	&	6.0	&	2.5	&	6.4\\
x01	&	PHOT, be?, re	&	-3.8:	&	0.4^a	&	2.0^b	&	2.0^b	&	-2.0	&	2.9	&	0.4	&	4.1\\
x02a	&	PHOT, asy	&	-10.1	&	2.6^a	&	-1.8	&	1.5	&	-5.4	&	1.3	&	-2.1	&	2.2\\
x02b	&	PHOT, asy, re	&	-12.5	&	-1.7	&	1.0^b	&	1.0^b	&	-4.3	&	-0.0	&	-1.5	&	0.9\\
x02c	&	PHOT		&	-1.5:	&	-3.3^a	&	1.5^b	&	1.5^b	&	0.9^b	&	0.9^b	&	0.6	&	2.0\\
x03	&	PHOT, RE	&	-3.5	&	1.0	&	1.4^d	&	-2.1^d	&	-7.2	&	-0.1	&	0.0	&	-2.1\\
	\hline
\multicolumn{10}{c}{47\,Tuc}	\\
x01	&	MOL		&	...	&	...	&	0.7:	&	2.8	&	0.8:^b	&	0.8:^b	&	2.0:	&	0.0:\\
x02	&	PHOT, asy	&	-8.8	&	-0.9	&	-0.9:	&	1.2:	&	-6.7	&	-0.1^a	&	-1.0:	&	0.5:\\
x03	&	PHOT, narrow	&	-1.0	&	4.4	&	4.4^d	&	0.4^d	&	-1.9	&	4.2	&	0.0:	&	3.0:\\
x04	&	PHOT, asy	&	-6.5:	&	-1.9^a	&	-4.1	&	1.5	&	-6.2	&	2.7	&	-7.0:	&	0.5:\\
x05	&	PHOT, mol	&	-4.8	&	4.0^a	&	-1.0	&	1.7	&	-2.6	&	1.8	&	-1.5:	&	1.0:\\
x06	&	Noisy continuum	&	...	&	...	&	...	&	...	&	-0.5:	&	1.0:^a	&	0.0:	&	1.0:^a\\
x07	&	Noisy continuum	&	...	&	...	&	6.0:^d	&	0.1:^d	&	4.7^b	&	4.7^b	&	-1.0:	&	1.0:^a\\
x08	&	BE, RE		&	-5.0:	&	-2.0:^a	&	1.0:	&	3.0:^a	&	0.0	&	1.5:^a	&	0.3	&	1.0:^a\\
x09	&	PHOT, asy	&	-7.8	&	-0.9^a	&	-4.6	&	-0.3	&	-7.9	&	-1.2	&	-8.0:	&	-1.5:^a\\
	\hline
\multicolumn{10}{c}{NGC\,6388}	\\
o01	&	PHOT, mol	&	-6.7	&	0.6	&	-0.4	&	1.6	&	-3.9	&	1.6^a	&	-1.3	&	1.4	\\
o02	&	phot		&	4.7	&	9.5	&	8.5^d	&	-0.4:^d	&	 3.1	&	7.4	&	7.0	&	1.4:^d	\\
o03	&	PHOT		&	-4.4	&	0.9	&	-1.7	&	1.2	&	-0.6^b	&	-0.6^b	&	-2.7	&	0.5	\\
o04	&	phot		&	-2.3:	&	0.1	&	0.1^b	&	0.1^b	&	-2.8	&	-0.6^a	&	-1.6	&	-0.1	\\
o05	&	PHOT, asy	&	-9.7	&	2.6^a	&	1.0^b	&	1.0^b	&	-2.3	&	0.6	&	-1.8	&	1.0	\\
o06	&	PHOT, mol	&	-6.7	&	-0.5	&	-0.1^b	&	-0.1^b	&	-5.0	&	-0.3^a	&	-3.3	&	-0.7	\\
o07	&	phot, mol, asy	&	-5.7	&	1.9	&	-0.6	&	1.0	&	-3.5	&	-0.9	&	-2.3	&	-0.3:	\\
o08	&	phot, mol	&	-2.3	&	0.6	&	-0.5	&	1.8	&	-1.0	&	0.8	&	-0.2	&	1.2	\\
o09	&	PHOT, wide	&	-0.0:	&	2.0	&	1.7^d	&	-0.1^d	&	1.1^b	&	1.1^b	&	1.2^b	&	1.3^b	\\
o10	&	PHOT, asy	&	-7.4	&	-2.6^a	&	4.5^d	&	2.1:^d	&	-0.2	&	3.3	&	1.9^b	&	1.9^b	\\
o11	&	PHOT, be?	&	1.0^b	&	1.0^b	&	0.7^b	&	0.7^b	&	-0.1^b	&	-0.1^b	&	0.8^b	&	0.8^b	\\
o12	&	PHOT, mol	&	-6.3	&	-0.6	&	0.0	&	1.6	&	-2.6:	&	-0.8^a	&	-2.5	&	0.7	\\
x01	&	phot, BE, RE	&	-0.9	&	1.9	&	4.1^d	&	0.4:^d	&	3.7^d	&	-3.4^d	&	3.3	&	-1.6^d	\\
x02	&	PHOT		&	-3.7	&	2.6	&	0.2^b	&	0.2^b	&	-1.8	&	0.9	&	-0.8	&	1.2	\\
x03	&	BE		&	24.3^c	&	19.4^c	&	17.0^d	&	1.9:^d	&	11.7^d	&	-4.7:^d&	14.7^d	&	-0.2:^d\\
x04	&	phot, mol	&	-2.3:	&	2.9	&	0.4^b	&	0.4^b	&	0.5^b	&	0.5^b	&	0.6^b	&	0.6^b	\\
x05	&	PHOT, mol, asy	&	-3.3	&	6.0^a	&	-0.5	&	2.8	&	-3.3	&	1.7	&	-1.2	&	3.3	\\
x06	&	MOL		&	1.0:^b	&	1.0:^b	&	2.8:^d	&	-0.4:^d	&	-2.6	&	0.3	&	0.5^b	&	0.5^b	\\
	\hline
\multicolumn{10}{c}{M\,15}	\\
x01	&	PHOT		&	-4.2	&	3.3^a	&	0.1^d	&	-1.7^d	&	0.0^d	&	-3.6^d	&	-0.1^b	&	0.1^b\\
x02	&	PHOT, be, re?	&	-12.9	&	-0.5^a	&	1.2	&	2.7	&	0.8	&	2.4	&	-1.7	&	3.9	\\
	\hline
\multicolumn{10}{c}{$\omega$\,Cen}	\\
x01	&	BE, re		&	-3.0^c	&	-4.0^c	&	+1.0^c	&	+0.5^c	&	+0.5^c	&	-1.8^c	&	0.0^c	&	-4.0^c\\
	\hline
\multicolumn{10}{p{0.73\textwidth}}{\small Positive values denote red-shifts (in km s$^{-1}$). Colons denote ill-defined values due to noise. Abbreviations: phot -- photospheric profile visible; asy -- asymmetric; mol -- molecular bands; be/re -- blue/red emission wing; capitalisation indicates strong features. Notes: $^a$ kink not present, value given near continuum level; $^b$ no significant velocity shift in bisector, approximate average given; $^c$ core at a greater red-shift than kink or no kink, due to an emission line or ($^d$) unusual profile. \normalsize} \\
	\hline
	\\ \\ \\ \\ \\ \\ \\
\end{tabular}
\end{center}
\end{table*}


\section{A simple model for estimating mass loss}

\subsection{The model}

\subsubsection{General form}

In our attempt to quantify mass-loss rates from our sample of stars, we have created a simplistic model, based on some crude approximations, which we use to model the H$\alpha$ line profile. Estimations of mass-loss rates based solely on line profiles are known to be notoriously uncertain, in part due to their inherent degeneracy in parameters such as velocity and turbulence (see e.g.\ Hagen 1978; Hagen et al.\ 1983). We address these problems in Section \ref{LimSect}.

\begin{figure}[!btp]
\resizebox{\hsize}{!}{\includegraphics{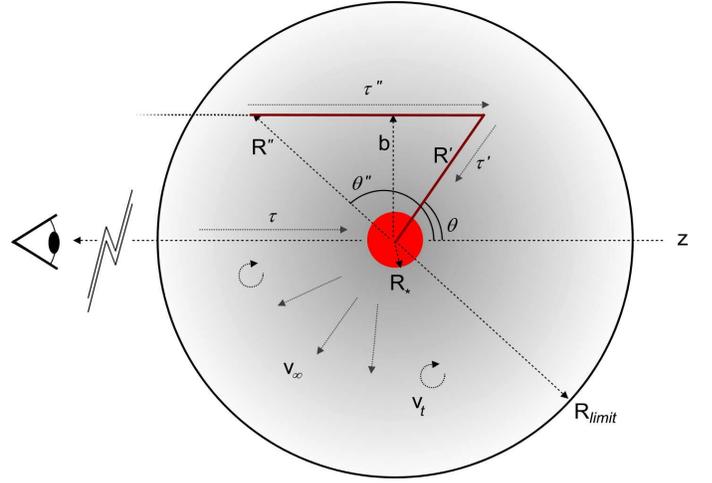}}
\caption[]{Illustration of our wind model, defining the symbols used in the text.}
\label{ModelFig}
\end{figure}

As illustrated in Fig.\ \ref{ModelFig}, we assume a spherically-symmetric wind expanding with constant speed $v_{\infty}$, emanating from a non-rotating star. We assume that the emission from the surface of the star in the region of the H$\alpha$ line can be simplified to a flat continuum, modified by a Gaussian-shaped absorption line, centred on H$\alpha$. This line profile is then modified by the stellar wind and finally convolved with a Kurucz stellar model with the same metallicity, temperature and gravity as the star, broadened by a macroturbulence parameter $v_{\rm turb}$, but without any hydrogen lines. Calculating the model in this way, rather than starting with the {\sc atlas9} model as the output from the photosphere allows an analytical representation of the photospheric emission which greatly improves calculation speed.

The width of the photospheric H$\alpha$ line is given by:
\begin{eqnarray}
	w &= \sqrt{ v_{\rm rms}^2 + v_{\rm turb}^2 }
				&= \sqrt{ \frac{3 k T_{\rm ph}}{\mu_{\rm H}} + v_{\rm turb}^2(r=R_{\ast}) } ,
\label{MdotwphotEqn}
\end{eqnarray}
where $k$ is Boltzmann's constant, $T_{\rm ph}$ is the effective temperature of the photosphere, $\mu_{\rm H}$ is the (reduced) atomic weight of hydrogen and $v_{\rm rms}$ is the root mean square velocity due to thermal motions (additional contributions to motion in the photosphere are covered by the `turbulence' term, $v_{\rm turb}$). Thus, the flux subtracted from the continuum by the photosphere, in terms of the continuum flux, can be described as simply:
\begin{equation}
	\left(\frac{A_{\rm ph}}{F_{\rm ctm}}\right)_v = \left( 1 - e^{-\tau_{\rm ph}} \right) \exp\left(\frac{-v^2}{2 w^2}\right) ,
\label{MdotFphotEqn}
\end{equation}
where $v$ is the radial velocity relative to H$\alpha$ and $\tau_{\rm ph}$ is the optical depth of the photospheric absorption in the line centre.

When calculating the wind contribution, we only consider the absorption and emission from bound-bound transitions for simplicity -- unless there is a chromosphere or other temperature enhancement (discussed later in this section), the temperatures in cool giant winds are never high enough to create sufficient amounts of free electrons that other contributions to opacity and line broadening become important. Thus, the observed flux from the star in velocity interval $dv$ can be modelled as:
\begin{eqnarray}
	\frac{F(v) dv}{F_{\rm ctm} dv} \!\! &= 1 - \displaystyle \frac{A_{\rm ph}(v) dv}{F_{\rm ctm} dv} + \frac{E_{\rm wind}(v) dv}{F_{\rm ctm} dv} - \frac{A_{\rm wind}(v) dv}{F_{\rm ctm} dv} ,
\label{MdotFEqn}
\end{eqnarray}
where $F_{\rm ctm}$ is the continuum emission near H$\alpha$; $E_{\rm wind}$ is the emission component of the wind, due to photons emitted by spontaneous emission directed \emph{into} the line of sight; and $A_{\rm wind}$ is the absorption component of the wind, due to photons absorbed \emph{from} the line of sight radiation from the stellar surface.

\subsubsection{The absorption term}

The absorption from the column in front of the star cannot be adequately modelled by a simple column of absorbing material, but must take into account the spherical nature of the star. Thus, we define an impact parameter, $b$, which represents the physical distance in the plane of the sky from the centre of the star.

We define the number density of absorbers (i.e.\ hydrogen atoms in the $n = 2$ level for Balmer $\alpha$) at a given point in the wind as being $n_2(r)$. This can be found in local thermodynamic equilibrium (LTE) through the Saha-Boltzmann equations (the validity of an LTE environment is discussed later). Using this, we can calculate the optical depth \emph{integrated over all velocities} in a particular radius bin $dr$ as being $d\tau(r) = n_2(r) \sigma dr$ for the velocity-integrated absorption cross-section $\sigma$. Here, we use:
\begin{eqnarray}
	\sigma = \frac{\pi e^2}{m c^2} f ,
\label{MdotSigmaEqn}
\end{eqnarray}
giving a value of $6.537 \times 10^{-12}$ m$^2$ using Gray (1992).

Integrating in annuli ($b$) over the observed stellar surface and outwards in radii, we can define the integrated optical depth of the absorption column at velocity $v$ as:
\begin{eqnarray}
	\tau(v) =& \displaystyle \int \limits_{r=R_{\ast}}^{\infty}  \int\limits_{b=0}^{R_{\ast}} \frac{1}{\sqrt{2\pi} w(r)} \exp \left( \frac{-(v+{v_{\infty} R_{\ast}} / b)^2}{2 w^2(r)} \right)
		\nonumber \\
	&	\displaystyle	\times \frac{2 \pi b db}{\pi R_{\ast}^2}  d\tau(r) dr ,
\label{MdotTauAbsnEqn}
\end{eqnarray}
with $v_{\infty} R_{\ast} / b$ being the projected velocity of the wind into our line of sight. We can thus define the absorption component in the column in front of the star as:
\begin{equation}
	\left(\frac{A_{\rm wind}}{F_{\rm ctm}}\right)_v = \left(1 - \frac{A_{\rm ph}(v)}{F_{\rm ctm}} \right) \left( 1 - e^{-\tau(v)} \right) .
\label{MdotFaEqn}
\end{equation}
%


\subsubsection{The emission term}

Assuming that emission in the H$\alpha$ line comes entirely from spontaneous emission, thus presently neglecting any chromosphere, we can describe the emission contribution at a particular velocity as:
\begin{equation}
	\left(\frac{E_{\rm wind}}{F_{\rm ctm}}\right)_v = \frac{\int\limits_{r=R_{\ast}}^{\infty} p_{\rm spon} f_{\rm spon} dr}{P_{\rm ctm}} ,
\label{MdotFeEqn}
\end{equation}
where $p_{\rm spon}$ is the power emitted from a shell of unit thickness \emph{integrated over all velocities}, $f_{\rm spon}$ is the fraction of that shell at observed radial velocity $v$ and $P_{\rm ctm}$ is the power emitted from the stellar continuum (for normalisation). This latter term is approximated as a blackbody, thus $P_{\rm ctm}$ is given by the familiar:
\begin{equation}
	P_{\rm ctm} \approx \frac{2 h \nu^3}{c^2} \frac{1}{\exp \left( \frac{h \nu}{k T_{\rm ph}} \right) - 1} \times 2 \pi \times 4 \pi R_{\ast}^2 d\nu,
\label{MdotPctmEqn}
\end{equation}
and the power $p_{\rm spon}$ is given as:
\begin{equation}
	p_{\rm spon} = 4 \pi r^2 \ h\nu n_3(r) A_{32} ,
\label{MdotPsponEqn}
\end{equation}
where $A_{32}$ is the Einstein co-efficient for the H$\alpha$ de-excitation (4.39 $\times 10^7$ s$^{-1}$ -- Menzel \& Pekeris 1935).

The fraction $f_{\rm spon}$ is slightly more complex, as the wind cannot be considered to be optically thin in all cases. We must account for the absorption of the emitted light by the wind between the point of emission and the observer, and it is computationally convenient to merge this with the $f_{\rm spon}$ term. We thus consider annuli of constant observed radial velocity (which in a non-rotating star also have a common depth in the wind from the observer's point of view), giving:
\begin{eqnarray}
	f_{\rm spon} =& \displaystyle	\int\limits_{\theta=\theta_{\ast}}^{\pi} \frac{1}{\sqrt{2\pi} w(r)} \exp \left( \frac{-(v - v_{\infty} \cos \theta)^2}{2 w^2(r)} \right) dv
				\nonumber \\
			&	\displaystyle	e^{-\tau_r''} \frac{2 \pi r^2 \sin\theta}{4 \pi r^2} d\theta,
\label{MdotFsponEqn}
\end{eqnarray}
where the Gaussian terms describe the fraction of atoms travelling in velocity $dv$, and the final term describes the fraction of the shell in that annulus. The lower integration limit (set to exclude flux emitted behind the star) is given as $\theta_{\ast} = \tan^{-1} (r/R_{\ast})$. The term $e^{-\tau_r''}$ describes the subsequent absorption of this emission, as mentioned above. Defining an optical path, $z$, from the emitting region towards the observer, we can calculate $\tau_r''$ (see Fig.\ 7):
\begin{equation}
	\tau_r'' = \int\limits_{z=0}^{\infty} n_2(r'') g_v'' \sigma dz ,
\label{MdotTauddEqn}
\end{equation}
where the distance from the stellar centre, $r''$ is given as:
\begin{equation}
	r'' = \sqrt{(z - r' \cos \theta)^2 + (r' \sin \theta)^2} ,
\label{MdotRddEqn}
\end{equation}
where $r'$ is the distance from the stellar centre to the emitting material. The function $g_v''$ selects the appropriate fraction of atoms in $dz$ at that velocity, namely:
\begin{equation}
	g_v'' = \frac{1}{\sqrt{2\pi} w(r'')} \exp \left( \frac{-v''^2}{2 w^2(r'')} \right),
\label{MdotGddEqn}
\end{equation}
with $v''$ defined as:
\begin{equation}
	v'' = v \cos \theta - v_{\infty} \cos \theta'' ,
\label{MdotVddEqn}
\end{equation}
where the angle $\theta''$ is given by:
\begin{equation}
	r'' sin \theta'' = r \sin \theta .
\label{MdotThetaddEqn}
\end{equation}

Using the above formulae, we can formulate a model where the only inputs are $v_{\infty}$, $T_{\rm ph}$, $\dot{M}$, $R_{\ast}$, $\tau_{\rm ph}$, $v_{\rm turb}(r=R_{\ast})$, and the radial dependences of temperature, density and (optionally) turbulence. The radial dependence of temperature can be considered to be constant for an isothermal wind and proportional to $r^{-2}$ for an adiabatic, iso-kinetic wind. The radial dependence of density must be proportional to $r^{-2}$ for a constant outflow velocity, from conservation of mass. $R_{\ast}$ and $T_{\rm ph}$ have been determined earlier from photometry and spectroscopy (Table \ref{ParamTable}).


\subsection{Limitations}
\label{LimSect}

Our model suffers from several limitations. The model assumes an LTE approximation, whilst non-LTE effects may be anticipated due to strong shocks and/or hot chromospheres. Attempts were made to model chromospheric heating in LTE using a wide range of temperature profiles up to $T \approx 7000$ K, but the model was unable to reproduce the emission seen in many of the line profiles. Also, free-bound recombinations were not taken into account, which would alter the $n=2$ and $n=3$ level populations. Although this is not significant at temperatures up to T $\sim$ 5500 K, it will play a r\^ole at higher temperatures where the plasma is ionised to a greater degree and is photo-electrically controlled (c.f.\ Thomas 1957; Dupree et al.\ 1984). We thus \emph{do not expect the model to accurately reproduce the emission seen in the line wings}.

Furthermore, although it contains some turbulence, our model assumes that the wind is isotropic. There is good evidence from both theory and observation (e.g.\ Asida \& Tuchman 1995; Tuthill et al. 1999) that red giant winds should not be isotropic. Magnetic effects in the warmer stars, and pulsation and shocking effects in the cooler stars will mean that density enhancements are likely to exist in the wind as well, which may play an important part in shaping the observed line profile.

Additionally, there are numerical limitations to consider: a limiting radius, $R_{\rm lim}$ must be taken for integration and a suitable step size must be chosen for $dv$, $dr$, $dz$, $d\theta$ and $db$. In our implementation, we use a constant step-size in all three parameters, for ease of computation. In terms of $R_{\rm lim}$, we recover the approximate true mass loss (assuming an inverse-square law for density and an isothermal, constant-velocity wind) by taking:
\begin{eqnarray}
	\dot{M}_{\rm true} &=& \displaystyle \frac{4}{3} {\dot{M}_{\rm computed}}  {\int\limits_{r=R_{\ast}}^{R_{\rm lim}} r^{-2} dr}   \left /  {\int\limits_{r=R_{\ast}}^{\infty} r^{-2} dr} \right.	\nonumber	\\
			&=& \displaystyle \frac{4}{3} {\dot{M}_{\rm computed}} \frac{(R_{\rm lim} / R_{\ast}) - 1}{R_{\rm lim} / R_{\ast}} ,
\label{MdotTrueEqn}
\end{eqnarray}
where the factor $\frac{4}{3}$ comes from the correction from a hydrogen mass-loss rate to a total mass-loss rate, assuming the wind is 75\% atomic hydrogen by mass. This last value is subject to error, given possible helium enrichment and/or molecular hydrogen.

We must also consider errors in the determination of the input values $T_{\rm ph}$ and $R_{\ast}$. Varying the temperature by 100 K can alter the mass-loss rate by approximately a factor of two, due to the sensitive variation of the excited populations with temperature -- the effects of a chromosphere are discussed later in this section. Similarly, the mass-loss rates we list are dependent on column densities in front of the star, thus they vary with the square of the stellar radius, and this then adds a $\sim$20\% uncertainty.

Finally, there is also considerable degeneracy among the parameters in the model, particularly between $\dot{M}$ and $v_{\infty}$, with a range of models able to produce similar spectra. Thus we conclude that our mass-loss rates are only reliable to an order of magnitude, and our wind and macroturbulence velocities ($v_{\rm turb}$) to within a few km s$^{-1}$. Errors could be derived from Monte-Carlo analysis of the modelling procedure, but this is prohibitively computationally expensive. Using this model, we cannot claim to correctly determine mass-loss rates below about 10$^{-8}$ M$_{\odot}$ yr$^{-1}$, depending predominantly on the stellar temperature (high temperatures have higher populations in $n = 2$ and 3).


\subsection{The fitting method and results}
\label{FitMethSect}

The observed spectrum of H$\alpha$ is also affected by lines of other elements, notably the neutral cobalt line at 6563.403 \AA\ and the singly-ionised hafnium line at 6563.860 \AA. Thus, we took the best-fit Kurucz spectrum (as described in Section \ref{ParamsSection}), convolved it with a macroturbulent velocity (the value of this velocity is discussed later) and resampled it to the wavelengths of the model bins. The data were also resampled to the model wavelengths, re-normalised in flux using the velocity ranges between 55 and 70 km s$^{-1}$ either side of the H$\alpha$ rest wavelength for some stars (denoted $a$ in Table \ref{ModelResultsTable}), or re-normalised by setting the integrated flux of the model and normalised data to be equal (denoted $b$). A goodness-of-fit parameter was then calculated. To better fit the core of the line profiles, in which blue-shifted absorption is the most reliable tracer of mass loss, we decided upon a chi-squared statistic, with a weight proportional to the inverse square of the observed flux. Method $b$ produces better fits to the H$\alpha$ profiles of stars with moderately-strong molecular lines, though we chose to accept the model with the minimum value of the chi-squared statistic.

Bearing in mind the above limitations, we have fitted models to all our targets, except those with the most extreme emission. This was done using a downhill-simplex search in four dimensions, using the parameters $\log(\dot{M})$, $\log(v_{\infty})$, $\log(\tau_{\rm ph})$ and $\log(v_{\rm turb})$. The first vertex of the initial simplex had values of $-6.5$, 1, 0 and 0.5, repectively; the other four vertices had these parameters changed in turn to --5.5, 1.33, --0.2 and 1.0, respectively, with the other parameters left at the original values. We also set $dv = 4$ km s$^{-1}$, $R_{\rm lim} = 4 R_{\ast}$, $dr = 0.035 R_{\ast}$, $dz = 0.07 R_{\ast}$, $d\theta = \pi / 25$ and $db = R_{\ast} / 25$. Fits were also attempted using the following parameters, which have smaller step sizes: $dv = 2$ km s$^{-1}$, $R_{\rm lim} = 8 R_{\ast}$, $dr = 0.02 R_{\ast}$, $dz = 0.04 R_{\ast}$, $d\theta = \pi / 40$ and $db = R_{\ast} / 25$, and a larger value for $R_{\rm lim}$, but they could not reproduce the observations as well in the majority of cases. We suspect that this is due to the poorly-modelled cobalt line, which does not appear in most cases to be as strong as the Kurucz spectrum would suggest, even when plausable forms of line broadening are taken into account. A few individual cases required special starting conditions in order to reproduce a good fit to the data, notably NGC\,362\,o10, which appears to have a net inflow.

In all cases, we chose an isothermal wind. The mass-loss rates required by winds with radially-decreasing temperatures were unreasonably high if the winds cooled by much more than $\sim$1000 K by 4 $R_{\ast}$. The macroturbulent velocity was not varied throughout the wind.

\begin{figure*}[!pbt]
\begin{center}
\resizebox{1.0\textwidth}{!}{\includegraphics{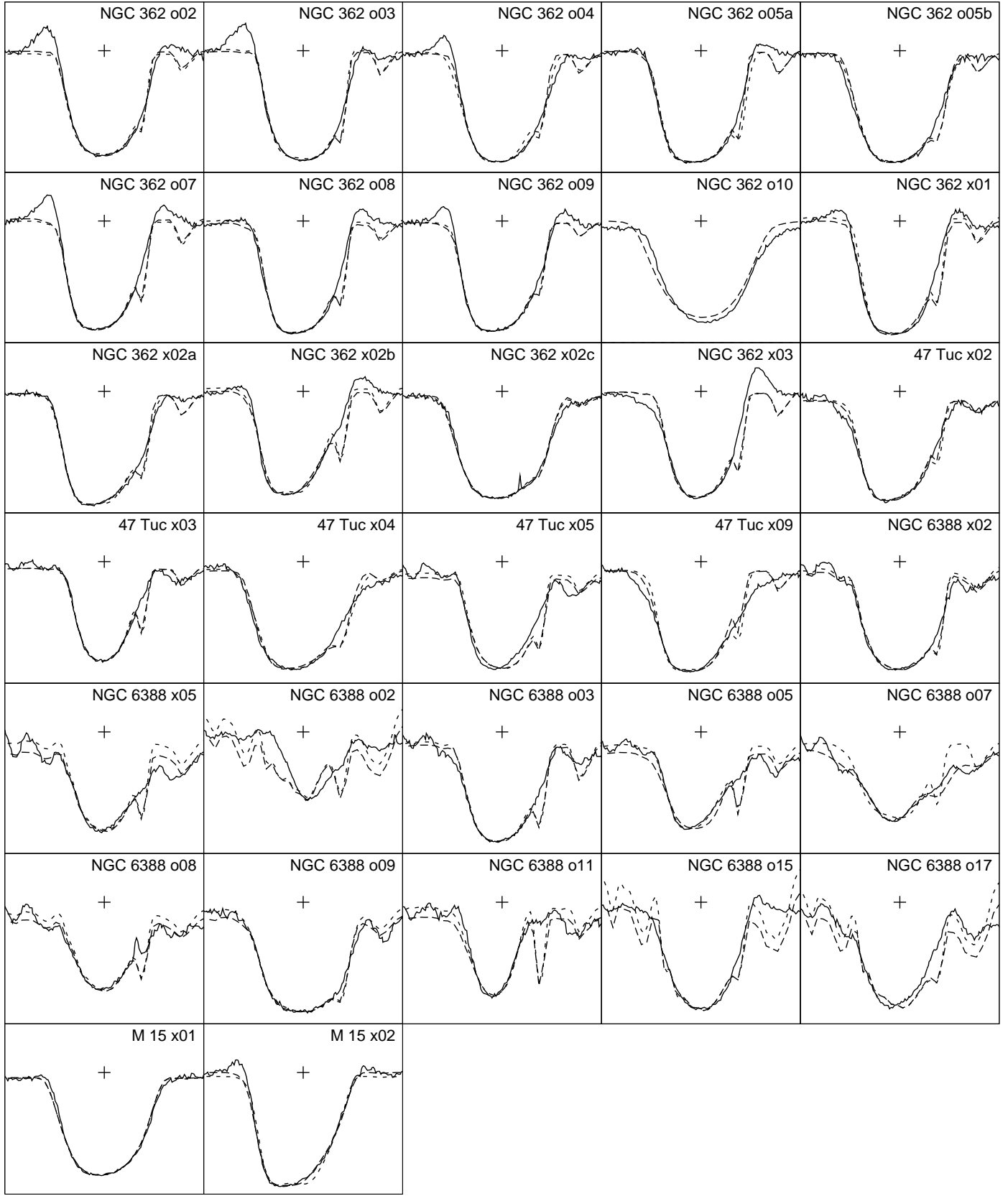}}
\caption[]{Results of model fitting to the H$\alpha$ lines. X-axis is velocity from --70 to +70 km s$^{-1}$, y-axis is normalised flux from zero to 1.4 times the continuum. The solid lines show the UVES spectrum; the long-dashed lines show the model fit using method (a); and the short-dashed lines show method (b). For NGC\,362\,o10, only method (a) was calculated, and this was done for an \emph{inflowing} wind. The cross represents zero velocity and the continuum flux level. The profiles are discussed in the text.}
\label{ModelsFig}
\end{center}
\end{figure*}

Fig.\ \ref{ModelsFig} and Table \ref{ModelResultsTable} show the results of the model fitting. Several of the cooler stars could not be fit due to strong interference from molecular lines, though the wind signature is often seen in their H$\alpha$ line profiles. Additionally, NGC\,362\,o01 and o06, 47\,Tuc\,x08, NGC\,6388\,x03 and $\omega$\,Cen\,x01 were not correctly modelled due to their strong emission.

In general, the absorption profiles otherwise match the observations reasonably well. The exception being the strength of the other lines in the profiles. In particular, the cobalt line (at +27 km s$^{-1}$ with respect to H$\alpha$) is much weaker in all the observed profiles (except the M\,15 stars) than the {\sc atlas9} model predicts. The same is true to a lesser extent of the hafnium line (at +48 km s$^{-1}$ with respect to H$\alpha$). Furthermore, it is interesting to note the variations in the observed line strengths when compared to the {\sc atlas9} model in NGC\,6388. While much of this is likely attributable to our relatively coarse temperature grid (250 K spacing), it is also possible that some temperatures have been inaccurately determined, or that the stars have substantially non-solar relative abundances. Detailed modelling of the photospheric spectra with {\sc phoenix} (Hauschildt et al.\ 1997; Baron \& Hauschildt 1998; Hauschildt et al.\ 2001) will be the subject of a forthcoming paper.

Two cases are worth highlighting here. All of the stars for which we have computed profiles show a significant amount of mass loss, with the exception of NGC\,362\,o10 and M\,15\,x01.

NGC\,362\,o10 cannot be modelled with an expanding wind, and has a H$\alpha$ line core that is red-shifted beyond any error in radial velocity. We find it to be accreting material at a substantial rate. It is not known to pulsate strongly, nor does it exhibit a significantly red (J--K) colour, but it may be that it has only recently started to pulsate and we observe it in the contracting phase of its pulsation cycle where mass may have the ability to flow inwards. Material ejected at a velocity of $\sim$10 km s$^{-1}$ during half of the pulsation cycle will reach a height of $< 10$ R$_{\ast}$. The escape velocity of NGC\,362\,o10 is one of the highest in our sample, and will drop to $< 10$ km s$^{-1}$ only at a distance of $\sim$76 R$_{\ast}$. If the driving mechanism responsible for removing the ejected material from the system is too weak in this star, then the ejected material is bound to fall back. Indeed, pulsation-related radial velocity variations of this order have been seen both in field giants (Lebzelter 1999) and globular cluster giants (Lebzelter et al.\ 2005), with wind models suggesting that inflowing material in the wind is also possible (Lebzelter et al.\ 2002). However, it remains a possibility that some of the blue wing of the H$\alpha$ line is filled in by emission, moving the absorption core redwards. A similar phenomenon may also be present in NGC\,6388\,o02, if the emission in this object is due to a chromosphere. However, given the high luminosity and low temperature of this object, it seems more likely that the emission seen here is due to pulsation.

Conversely, M\,15\,x01 appears to have both very little emission and very little (though still clearly non-zero) mass loss. It perhaps represents the most quiescent of our sample, which, given its low metallicity, is interesting. M\,15\,x02, on the other hand, displays both emission wings and a strong blue-shifted outflow signature, so the metallicity dependence is far from clear-cut.

\begin{table}[!thbp]
\begin{center}
\caption{Results of fitting the H$\alpha$ line profiles with the wind model described in Section 5.}
\label{ModelResultsTable}
\begin{tabular}{lr@{}l@{},rrrc}
    \hline \hline
ID     & \multicolumn{2}{c}{$\dot{M}$}  & v_\infty  &  $\tau_{\rm ph}$  & $v_{\rm turb}$ & $\chi^2_{\rm r}$ \  & * \\
       & \multicolumn{2}{c}{(M$_{\odot}$ yr$^{-1}$)}&  \multicolumn{1}{c}{\llap{(}km s$^{-1}$\rlap{)}}  &    & \llap{(}km s$^{-1}$\rlap{)} &   &  \\
    \hline
\multicolumn{8}{c}{NGC\,362} \\
o02 & 7.6&$\times 10^{-8}$: & >  8.4 & 1.93 &  9.2 &  240 & a \\
o03 & 1.5&$\times 10^{-7}$: & > 17.4 & 2.38 &  7.8 &  432 & a \\
o04 & 4.3&$\times 10^{-7}$: & >  6.7 & 1.96 & 10.3 &  386 & a \\
o05a& 1.8&$\times 10^{-6}$  &    7.2 & 1.60 & 13.0 &  604 & b \\
o05b& 1.3&$\times 10^{-6}$  &    5.9 & 1.53 & 15.2 &  495 & b \\
o07 & 5.7&$\times 10^{-6}$: & > 12.4 & 1.78 &  8.0 &  427 & a \\
o08 & 1.5&$\times 10^{-5}$  &   12.3 & 1.94 &  7.8 &  507 & a \\
o09 & 1.9&$\times 10^{-7}$: & > 12.9 & 1.74 &  9.1 &  233 & a \\
o10&--1.2&$\times 10^{-7}$  &   -5.9 & 1.12 & 19.3 &  167 & c \\
x01 & 3.7&$\times 10^{-6}$: & >  6.9 & 1.60 &  9.7 &  660 & b \\
x02a& 9.9&$\times 10^{-6}$  &   12.1 & 1.63 & 11.4 &  373 & a \\
x02b& 1.3&$\times 10^{-5}$  &   16.1 & 0.84 & 11.5 &  350 & a \\
x02c& 1.3&$\times 10^{-7}$  &   10.7 & 1.71 & 14.6 &   93 & a \\
x03 & 2.1&$\times 10^{-6}$: & > 11.3 & 1.89 &  7.1 &  499 & a \\
    \hline
\multicolumn{8}{c}{47\,Tuc} \\
x02 & 1.3&$\times 10^{-6}$  &   14.7 & 1.58 &  8.2 &  277 & a \\
x03 & 5.8&$\times 10^{-7}$  &    6.5 & 1.39 &  7.2 &  165 & a \\
x04 & 1.1&$\times 10^{-6}$  &   10.7 & 1.22 & 18.9 &  478 & b \\
x05 & 2.1&$\times 10^{-7}$  &   18.6 & 1.90 &  8.0 &  938 & a \\
x09 & 7.2&$\times 10^{-6}$  &   16.3 & 1.68 &  6.0 &  504 & a \\
    \hline
\multicolumn{8}{c}{NGC\,6388} \\
x02 & 3.5&$\times 10^{-7}$  &   18.5 & 2.04 & 10.1 &  412 & b \\
x05 & 2.4&$\times 10^{-7}$  &   14.2 & 1.38 &  8.3 &  446 & a \\
o02 & 1.3&$\times 10^{-6}$: &\gtrsim1.5&0.37& 14.3 &  420 & b \\
o03 & 1.4&$\times 10^{-6}$  &    9.5 & 1.72 &  8.5 &  598 & a \\
o05 & 4.6&$\times 10^{-6}$  &   10.6 & 0.55 & 12.6 &  293 & b \\
o07 & 1.4&$\times 10^{-7}$  &    5.3 & 0.01 & 34.0 &   92 & a \\
o08 & 3.5&$\times 10^{-6}$  &    7.4 & 0.84 & 10.7 &  190 & a \\
o09 & 7.9&$\times 10^{-6}$  &    7.0 & 1.99 & 12.3 &  407 & a \\
o10 & 3.6&$\times 10^{-6}$  &   10.5 & 0.41 &  0.7 &  489 & b \\
o11 & 2.0&$\times 10^{-7}$: &   15.4:& 1.81 &  8.6 & 1066 & a \\
o12 & 2.1&$\times 10^{-7}$: &   17.6:& 1.70 & 11.1 & 1112 & b \\
    \hline
\multicolumn{8}{c}{M\,15} \\
x01 & 4.3&$\times 10^{-8}$  &   10.0 & 1.60 & 14.0 &   45 & a \\
x02 & 6.4&$\times 10^{-7}$: & > 14.9 & 1.24 & 13.6 &  175 & a \\
    \hline
\multicolumn{8}{p{0.46\textwidth}}{\small Notes: The fits are to an isothermal, constant velocity wind, without the inclusion of a chromosphere. The value $\chi^2_{\rm r}$ is a \emph{relative} goodness-of-fit parameter only. The coolest stars did not yield any meaningful results due to strong interference from molecular bands (see text). Errors are also discussed in Section \ref{LimSect}. Values with a trailing colon are uncertain: these and lower limits to velocities are discussed in Section \ref{ModChromSect}. $^{\ast}$Method (a) is normalised using a continuum taken at $\pm$ (55 -- 70) km s$^{-1}$ from H$\alpha$; method (b) is normalised using the entire line from $-$70 to +70 km s$^{-1}$; method (c) uses the same normalisation as (a), but uses an inflowing wind (see text). The method used was chosen to give the best fit, i.e.\ lowest $\chi^2_{\rm r}$. \normalsize} \\
    \hline
\end{tabular}
\end{center}
\end{table}
\normalsize

\subsection{The influence of chromospheres on the spectral line profiles}
\label{ModChromSect}

As noted, our simple model does not include all the physics necessary for high-energy plasmas. Thus, we cannot expect to accurately model a chromosphere and reproduce the observed line profile. However, we have experimented with the introduction of a layer in the wind with variable temperature, density and outflow velocity. By inserting a near-stationary layer with the temperature and density enhanced relative to the wind, we can mimic a chromosphere and use it to qualitatively explore its effects.

The observed emission wings seen in several profiles cannot be reproduced by a localised density enhancement (with the corresponding change in velocity). We find that, in order to reproduce the emission wings, there needs to be a significant fraction of gas at a temperature well above that of the photosphere ($T \gtrsim 7000 K$), meaning it will be largely ionised -- i.e.\ a chromosphere. We find it is likely restricted to radii of $\lesssim 2 R_{\ast}$ for stars showing only a blue emission wing, as this is likely a result of the red wing being blocked by the stellar disk; whereas stars showing both red and blue emission wings are likely to also have a large amount of material at radii $\gtrsim 1.5 R_{\ast}$. It seems further likely that stars showing only a red emission wing, or P-Cygni-type profile, have the blue wing reabsorbed by the material in front of the emitting region.

The existence of a chromosphere can mean that the true (instantaneous) mass-loss rates differ significantly from the mass-loss rates we have calculated above. The existence of emission wings creates two competing effects: firstly, the temperature enhancement increases the fraction of atoms in the $n = 2$ level, leading us to overestimate the mass-loss rate; secondly, the infilling of the blue-shifted absorption core leads us to underestimate both the mass-loss rate and expansion velocity. In general our modelling suggests that the former term dominates, and in the stars showing the strongest wings, the mass-loss rate may be overestimated by up to a factor of roughly ten and the velocity underestimated by up to 10 km s$^{-1}$. Weaker emission wings will cause less of an effect. The affected mass-loss rates are listed as uncertain, and the velocities as lower limits, in Table \ref{ModelResultsTable}.


\section{Discussion}


\subsection{Spectroscopic correlations with infrared excess}

\begin{table}[!Htb]
\begin{center}
\caption{Properties of stars with and without IR excess in clusters NGC\,362 and NGC\,6388.}
\label{ExcessStatsTable}
\begin{tabular}{l@{\qquad\qquad}r@{\ $\pm$\,}r@{\ $\pm$\,}r@{\qquad\quad}r@{\ $\pm$\,}r@{\ $\pm$\,}r}
    \hline\hline
Property                                     & \multicolumn{6}{c}{Objects with:}   \\
                                             & \multicolumn{3}{c}{No excess (24)}    & \multicolumn{3}{c}{Excess (23)} \\
    \hline
Effective	&	3886	&	280	&	57	&	3893	&	327	&	103	\\
Temp. (K)		     \\
    \hline
    \multicolumn{7}{c}{H$\alpha$}\\
Core offset	&	--3.5	&	4.3	&	0.9	&	--0.4	&	9.1	&	2.9	\\
Kink offset	&	1.9	&	3.3	&	0.7	&	3.5	&	6.1	&	1.9	\\
Velocity range	&	--5.4	&	3.9	&	0.8	&	--3.8	&	5.2	&	1.6	\\
    \hline
    \multicolumn{7}{c}{Ca II 8498 \AA}\\
Core offset	&	1.5	&	2.2	&	0.4	&	2.7	&	5.3	&	1.7	\\
Kink offset	&	1.9	&	2.2	&	0.5	&	0.8	&	1.4	&	0.4	\\
Velocity range	&	--0.4	&	2.7	&	0.6	&	1.9	&	5.3	&	1.7	\\
    \hline
    \multicolumn{7}{c}{Ca II 8542 \AA}\\
Core offset	&	--0.7	&	2.2	&	0.5	&	--0.6	&	5.3	&	1.7	\\
Kink offset	&	2.2	&	2.9	&	0.6	&	0.0	&	2.3	&	0.7	\\
Velocity range	&	--2.8	&	2.6	&	0.5	&	--0.6	&	7.3	&	2.3	\\
    \hline
    \multicolumn{7}{c}{Ca II 8662 \AA}\\
Core offset	&	0.3	&	2.3	&	0.5	&	1.6	&	4.8	&	1.5	\\
Kink offset	&	2.0	&	1.9	&	0.4	&	1.0	&	2.0	&	0.6	\\
Velocity range	&	--1.7	&	2.3	&	0.5	&	0.6	&	5.8	&	1.8	\\
     \hline
    \multicolumn{7}{c}{Ca II Average}\\
Core offset	&	0.4	&	2.0	&	0.4	&	1.3	&	5.0	&	1.6	\\
Kink offset	&	2.0	&	2.1	&	0.4	&	0.6	&	1.6	&	0.5	\\
Velocity range	&	--1.6	&	2.1	&	0.4	&	0.6	&	5.9	&	1.9	\\
    \hline\hline
	                                     & \multicolumn{6}{c}{All objects in:}   \\
                                             & \multicolumn{3}{c}{NGC\,362 (12)}    & \multicolumn{3}{c}{NGC\,6388 (18)} \\
    \hline
Effective	&	4221	&	186	&	46	&	3713	&	219	&	52	\\
Temp. (K)		     \\
    \hline
    \multicolumn{7}{c}{H$\alpha$}\\
Core offset	&	--3.3	&	4.6	&	1.3	&	--1.7	&	7.4	&	1.8	\\
Kink offset	&	2.5	&	3.6	&	1.0	&	2.7	&	4.9	&	1.2	\\
Velocity range	&	--5.8	&	4.5	&	1.3	&	--4.4	&	3.9	&	0.9	\\
    \hline
    \multicolumn{7}{c}{Ca II 8498 \AA}\\
Core offset	&	1.9	&	1.3	&	0.4	&	2.1	&	4.5	&	1.0	\\
Kink offset	&	3.0	&	2.7	&	0.8	&	0.9	&	0.9	&	0.2	\\
Velocity range	&	--1.1	&	2.1	&	0.6	&	1.2	&	4.5	&	1.1	\\
    \hline
    \multicolumn{7}{c}{Ca II 8542 \AA}\\
Core offset	&	0.2	&	1.9	&	0.6	&	--0.5	&	3.8	&	0.9	\\
Kink offset	&	3.4	&	2.9	&	0.8	&	0.4	&	2.5	&	0.6	\\
Velocity range	&	--3.2	&	3.2	&	0.9	&	--0.9	&	5.2	&	1.2	\\
    \hline
    \multicolumn{7}{c}{Ca II 8662 \AA}\\
Core offset	&	0.9	&	1.3	&	0.4	&	0.7	&	4.3	&	1.0	\\
Kink offset	&	3.2	&	1.8	&	0.5	&	0.7	&	1.1	&	0.3	\\
Velocity range	&	--2.3	&	2.0	&	0.6	&	0.0	&	4.6	&	1.1	\\
     \hline
    \multicolumn{7}{c}{Ca II Average}\\
Core offset	&	1.0	&	1.2	&	0.3	&	0.7	&	4.1	&	1.0	\\
Kink offset	&	3.2	&	2.4	&	0.7	&	0.7	&	1.2	&	0.3	\\
Velocity range	&	--2.2	&	2.2	&	0.7	&	0.1	&	4.5	&	1.1	\\
    \hline\hline
\multicolumn{7}{p{0.45\textwidth}}{\small Notes: The other clusters have been excluded to reduce selection effects. The first error column gives the standard deviation from the mean of the sample, the second error is the error in the mean. Velocity range is defined by kink velocity (or redmost velocity when the kink is not present) minus core velocity. Units are km s$^{-1}$, except where noted. \normalsize} \\
    \hline
\end{tabular}
\end{center}
\end{table}

One of our original aims was to identify differences between the populations with and without mid-IR excess, with the expectation that the excesses were likely due to dust-rich winds. We present some statistical properties of our sample in Table \ref{ExcessStatsTable} (47\,Tuc has not been included for comparison due to its small sample size, and the strong molecular bands and effects of shocking in its cooler stars).

It appears as though there is little, if any, statistical difference between IR-normal and IR-excessive stars whatsoever. If we consider the entire sample, the temperature of the IR-excessive stars is slightly lower, but this appears to be due to selection effects when making up the original sample and disappears when we only consider the clusters where we have both excessive and non-excessive objects (as in Table \ref{ExcessStatsTable}). Interestingly, the temperatures of the two sets of stars are also identical, contrary to the expectation the IR-excessive stars should be cooler due to their advanced evolution (which leads to dust production and IR-excess).

One possibility for this is that the intrinsic brightness variability of the stars, due to their pulsation, is causing some objects to appear to have an IR excess when they do not. As the photometry used to assign stars with an excess is not strictly co-temporal, this suggests that it is normal variability, rather than dust, which separates our `IR excessive' stars from the rest of the sample. As such, we would expect there to be little or no difference between the parameters, as is the case. However, the IR excesses listed in OFFR02 are of order of a magnitude, while the K-band amplitudes found for these stars (Table \ref{ObjectTable}) are $\leq 0.6$ magnitudes, suggesting this cannot be the only reason. Hagen et al.\ (1983) find little correlation between mass-loss rate and dust-to-gas ratio, thus we should perhaps not expect any differences in the IR-excessive stars, which presumably contain more dust.

It should be noted that the objects with excess show higher variances in velocity offsets and ranges, particularly in the core velocity. Considering the period-magnitude diagram (Fig.\ \ref{UVESPL}) we find that the objects with longer periods are the objects with stronger pulsations (Table \ref{ObjectTable}). All the objects for which we have variability information are (barring NGC\,362\,o01) identified as IR-excessive in the {\sc isocam} photometry, as are the few most luminous and coolest stars (see Fig.\ \ref{UVESHRD}).

The chromospheric activity appears more pronounced in IR-normal stars, but this bias is most likely due to selection effects, since the stars in NGC\,362 (which contains most stars showing chromospheric activity) are hotter and more luminous, therefore are more likely to have chromosphere-driven winds, rather than pulsation-driven winds. However, if IR excess is linked with dust formation, then it would not be surprising that dust formation is inhibited while the hot chromosphere is still in place.

These factors suggest that those objects labelled IR-excessive stars are, by and large, more active objects with greater variability -- hence more evolved objects. Further evidence for this comes from Fig.\ \ref{UVESHRD}, which shows that the most luminous, coolest objects are all IR-excessive. We do not find any significant correlation between period and temperature, nor between core asymmetry and luminosity, nor core asymmetry and temperature. It is important to realise that the stars are within a factor $\sim$4 of the RGB tip: stars lower on the RGB are likely to pulsate less, and dust formation at lower luminosities may be more stochastic and a by-product of outflows, rather than an integral part of driving a wind.

Also in Table \ref{ExcessStatsTable}, we compare the two clusters, which are at significantly different metallicities (Table \ref{ClusterTable}). Given our relatively small sample size, there are no statistically significant variations between the clusters in the H$\alpha$ line, despite the clear difference in stellar temperature. This suggests that temperature is not a strong influence on the outflow velocity.

\begin{figure}[t!bp]
\resizebox{\hsize}{!}{\includegraphics[angle=270]{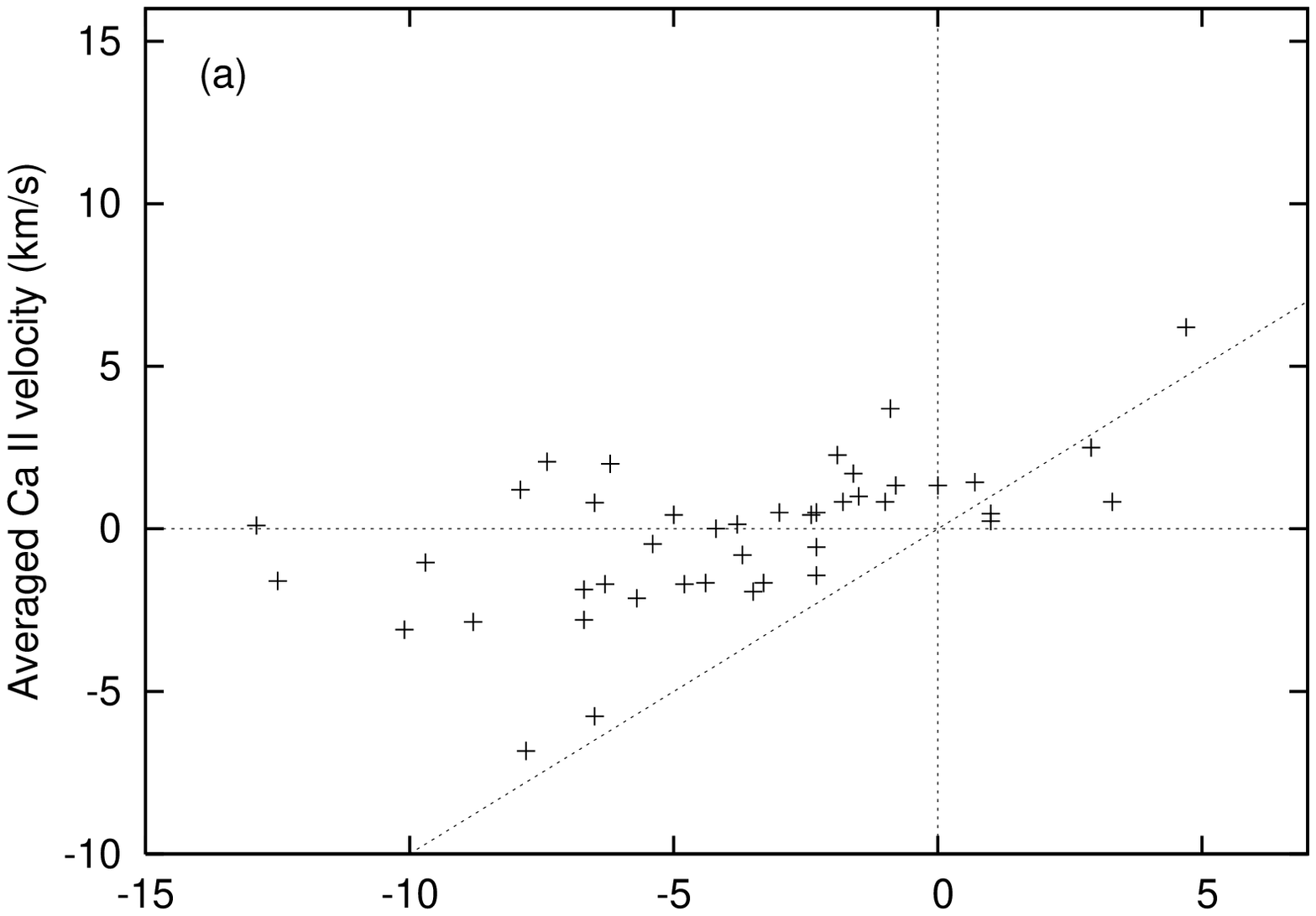}}
\resizebox{\hsize}{!}{\includegraphics[angle=270]{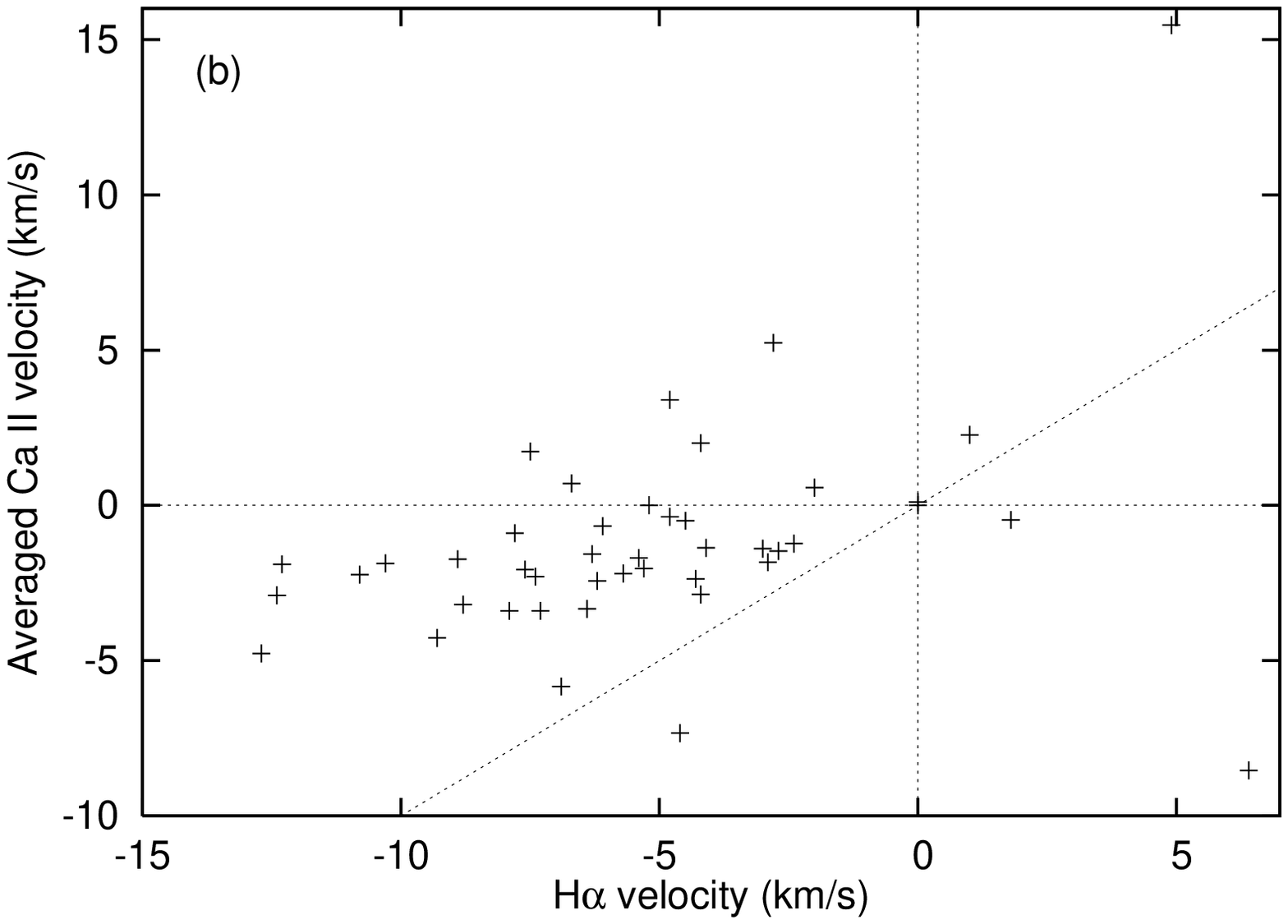}}
\caption[]{Plots of H$\alpha$ core velocity against calcium triplet core velocity (averaged over the three lines). Velocity is taken with respect to (a) the stellar radial velocity and (b) with respect to the line `kink'; see Sect.\ 4 for details. The dotted lines indicate zero and equal velocities for both the H$\alpha$ and Ca II lines.}
\label{HaCaVel}
\end{figure}


\subsection{Line core velocities}

Fig.\ \ref{HaCaVel} shows the variation of hydrogen versus calcium line core velocities, with respect to the stellar rest velocity and the `kink' velocity described previously. Clearly a correlation exists in both cases, although the calcium lines are only perturbed by around 30\% of the velocity of the hydrogen lines. This suggests that the lower layers of the atmosphere, where the calcium lines form, have not yet been accelerated to the terminal velocity. The two outliers showing redshifted H$\alpha$ cores in the lower panel of Fig.\ \ref{HaCaVel} are NGC\,362\,o01 and NGC\,6388\,x03, both of which show strong emission in H$\alpha$.

Though their results are mostly for the Na D lines rather than H$\alpha$ and Ca II, Bates et al.\ (1993) find that low-luminosity stars show only small core velocity shifts, though these increase beyond a luminosity of $\approx 1000$ L$_{\odot}$ and become more scattered. Most of our sample is above 1000 L$_{\odot}$, thus we cannot verify this claim. We do see considerable scatter, but do not find any correlation between line core velocity and luminosity within the luminosity range of our sample.


\subsection{Line profile shapes}

\begin{figure}[!tbp]
\resizebox{\hsize}{!}{\includegraphics{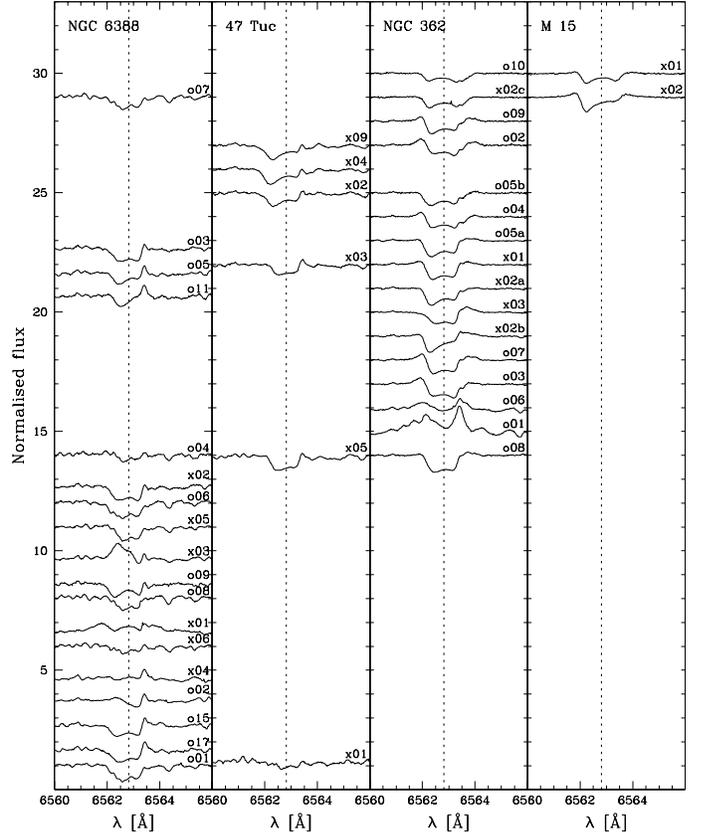}}
\caption[]{H$\alpha$ line profiles of the target stars, sorted by temperature (temperature increases upwards), with the appropriate {\sc atlas9} model spectrum subtracted. Spectra have been normalised to their continuum fluxes as in Figs.~\ref{SpectraFig} \& \ref{HalphaFig} and shifted to laboratory wavelengths. Spectra for 47\,Tuc\,x06, x07 and x08, and $\omega$\,Cen\,x01 have not been shown, as they are too cool to be accurately matched to an {\sc atlas9} model. The four panels show objects in different clusters, with metallicity decreasing from left to right. A Co {\sc i} line at 6563.4 \AA\ appears to have been oversubtracted in NGC\,6388, giving the false impression of red-shifted emission.}
\label{HalphaMinusKuruczFig}
\end{figure}

\begin{figure}[!tbp]
\resizebox{\hsize}{!}{\includegraphics{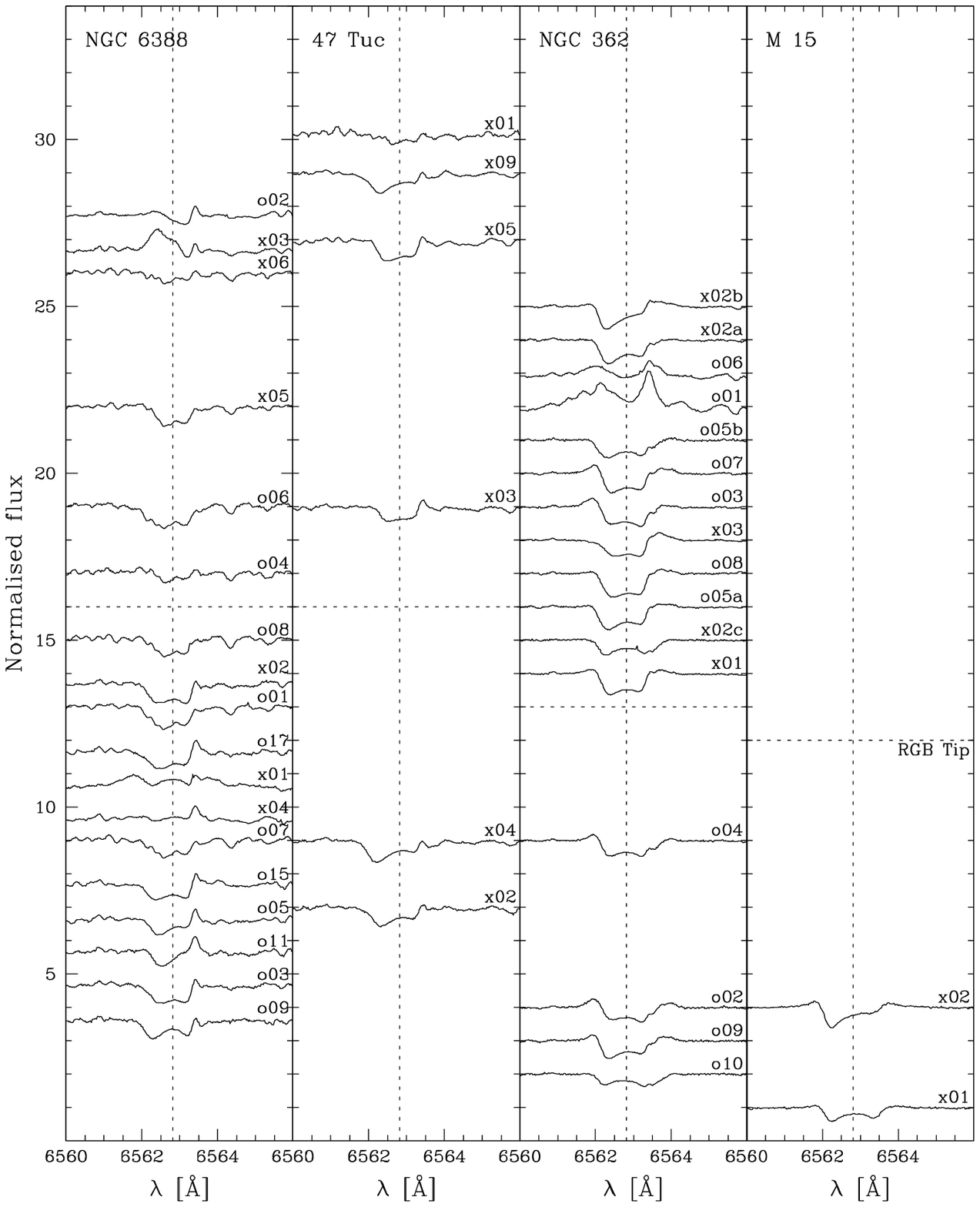}}
\caption[]{As Fig.~\ref{HalphaMinusKuruczFig} but with spectra sorted by luminosity (most luminous at top).}
\label{HalphaMinusKuruczFig2}
\end{figure}

Fig.\ \ref{HalphaFig} clearly shows unusually deep line profiles. To investigate this, and the other line profile features already noted, the {\sc atlas9} model corresponding to the temperature and metallicity of the star was subtracted from the H$\alpha$ profile. Interpolations were made to more accurately model the correct metallicity. These results are presented in Figs.\ \ref{HalphaMinusKuruczFig} \& \ref{HalphaMinusKuruczFig2}.

Two things are immediately apparent from these figures: firstly, that the absorption components are indeed much deeper than the {\sc atlas9} modelling would suggest; and secondly, that emission in both red and blue components persists. We have discussed a chromospheric model for producing the emission. The deep absorption could arguably be produced by the same phenomenon: from an optically thick chromosphere.

Another facet of Figs.\ \ref{HalphaMinusKuruczFig} \& \ref{HalphaMinusKuruczFig2} is the progression of emission and line asymmetry with spectral type and metallicity. Note first that in the figures, the emission spike present in most stars in NGC\,6388 (and possibly 47\,Tuc) at 6563.4 \AA\ co-incides with a Co {\sc i} line. This line has been grossly over-subtracted, suggesting that there may be a cobalt deficiency in these objects, or an error in our temperature calculations for intermediate ($\sim$4000 K) temperatures. However, examination of the original data (Fig.\ \ref{HalphaFig}) suggests that at least some part of this emission may be still associated with the H$\alpha$ line.

Looking within clusters, blue asymmetry and line depth appear to increase as luminosity increases towards 2400 L$_{\odot}$ (or as temperature decreases towards $\sim$3900 K). As luminosity continues to increase and temperature decrease, profiles become more variable. Similarly, broad emission wings only appear in stars below $\approx$ 2400 L$_{\odot}$ (above $\sim$3900 K), and in cool stars is only present if they show high photometric variability and pulsation shocks. The emission wings are only seen clearly in NGC\,362 and M\,15, the low-metallicity clusters (and possibly NGC\,6388\,o02, though at this luminosity and temperature this could be due to pulsation), with little or no emission seen in the higher-metallicity NGC\,6388 and 47\,Tuc. While our low-metallicity sample are generally warmer stars where chromospheres may be expected, and where there are fewer strong spectral lines to mask their signature, the complete lack of apparent line wing emission in the high-metallicity stars is striking. This may mean that mass loss from metal-poor stars is dominated by their chromospheres longer in the evolution of the star. Note that the deep H$\alpha$ line, blue asymmetry and emission appear to disappear for the cool stars (which are \emph{not} the most luminous) in NGC\,6388. We remind the reader that we cannot readily differentiate whether stars below $\approx$2400 L$_{\odot}$ are on the RGB or AGB, which could confuse these results.


\subsection{Mass-loss rates and wind velocities}

The calculated mass-loss rates can be found in Table \ref{MassLossTable}, which also includes the following semi-empirically-determined mass-loss rates from Reimers (1975), SC05 and Nieuwenhuijzen \& de Jager (1990, hereafter NdJ90), respectively (in M$_{\odot}$ yr$^{-1}$):
\begin{equation}
	\dot{M}_{\rm R75}     = 4 \times 10^{-13} \eta \frac{LR}{M} \\
\label{MdotEqun1}
\end{equation}
\begin{equation}
	\dot{M}_{\rm SC05}    = 8 \times 10^{-14} \frac{LR}{M}
		  		\left(\frac{T_{\rm eff}}{4000\mbox{K}}\right)^{3.5}
				\left(1 + \frac{g_\odot}{4300 g_\ast}\right) \\
\label{MdotEqun2}
\end{equation}
\begin{equation}
	\dot{M}_{\rm NdJ90}   = 9.631 \times 10^{-15} L^{1.42} M^{0.16} R^{0.81},
\label{MdotEqun3}
\end{equation}
for luminosity $L$, radius $R$, mass $M$ in solar units and stellar and solar gravities $g_\ast$ and $g_\odot$ (see Table \ref{ParamTable}, where $M =$ 0.8 M$_{\odot}$ has been adopted); and a semi-empirically determined constant $\eta$, which is usually in the range 0.4 $< \eta <$ 3 (Stancliffe \& Jeffery 2007), with $\eta \sim 0.5$ being most favoured (SC05). Although values up to $\eta = 10$ have been used effectively in specific cases (e.g.\ Straniero et al.\ 1997 in modelling for intermediate-mass field stars), whereas SC05 express $\eta$ in terms of $T_{\rm eff}$ and $g$, based on a heuristic model for the chromospheric driving of mass-loss. In this work, we assume $\eta =$ 1, as it provides a closer match to the SC05 results. We also compare our results to those of OFFR02, who determine mass-loss rates near-independently (save for identical photometry with which we have calculated the stellar parameters) and Gratton (1983), which is discussed later in this section. OFFR02 use dust masses inferred from infrared colours and scale them to estimate the total mass loss, via a gas-to-dust ratio, for which they adopt:
\begin{equation}
	M_{\rm gas} / M_{\rm dust} = 200 \times 10^{-({\rm [Fe/H]}+0.76)} .
\label{MgasMdustEqun}
\end{equation}

\begin{table}[!Htbp]
\begin{center}
\caption{Estimated mass-loss rates and equivalent widths (EW) of the H$\alpha$ line profile emission contribution.}
\label{MassLossTable}
\begin{tabular}{lc,@{\,},,,,,}

\hline\hline
ID    &   EW     & \multicolumn{6}{c}{$\dot{M} (10^{-7} \mbox{M}_{\odot} \mbox{yr}^{-1})$} \\[-2pt]
      &          & \multicolumn{6}{c}{$\overbrace{\qquad\qquad\qquad\qquad\qquad\qquad\qquad\qquad\qquad}$} \\[-3pt]
      &   (\AA)  & \multicolumn{1}{c}{(a)}&\multicolumn{1}{c}{(b)}&\multicolumn{1}{c}{(c)}&\multicolumn{1}{c}{(d)}&\multicolumn{1}{c}{(e)}&\multicolumn{1}{c}{(f)}\\
    \hline
\multicolumn{8}{c}{NGC\,362} \\
o01    &   \       &   \    &   1.3   &  1.1   &  0.26  &   \   &   \     \\[-0.00pt]
o02    &   0.14    &   0.8  &   0.3   &  0.15  &  0.05  &   \   &   3.2   \\[-0.00pt]
o03    &   0.14    &   1.5  &   1.2   &  1.0   &  0.24  &   \   &   5.8   \\[-0.00pt]
o04    &   0.10    &   4.3  &   0.6   &  0.34  &  0.10  &   \   &   3.5   \\[-0.00pt]
o05a   &   0.11    &   18   &   1.0   &  1.3   &  0.19  &   \   &   4.6   \\[-0.00pt]
o05b   &   \       &   13   &   1.4   &  0.99  &  0.29  &   \   &   \     \\[-0.00pt]
o06    &   \       &   \    &   1.4   &  1.2   &  0.28  &   \   &   \     \\[-0.00pt]
o07    &   0.18    &   57   &   1.2   &  0.98  &  0.24  &   \   &   6.1   \\[-0.00pt]
o08    &   0.13    &   154  &   1.1   &  0.88  &  0.22  &   \   &   5.2   \\[-0.00pt]
o09    &   0.11    &   1.9  &   0.15  &  0.07  &  0.02  &   \   &   2.2   \\[-0.00pt]
o10    &   \       &  -1.2  &   0.13  &  0.07  &  0.02  &   \   &   \     \\[-0.00pt]
x01    &   0.10    &   37   &   0.8   &  0.59  &  0.16  &   4.4 &   4.1   \\[-0.00pt]
x02a   &   \       &   99   &   1.4   &  0.98  &  0.29  &       &   \     \\[-0.00pt]
x02b   &   0.16    &   133  &   1.6   &  1.2   &  0.34  &   \   &   6.4   \\[-0.00pt]
x02c   &   \       &   1.3  &   0.8   &  0.48  &  0.16  &   2.8 &   \     \\[-0.00pt]
x03    &   0.16    &   21   &   1.1   &  0.85  &  0.21  &   3.6 &   5.4   \\
    \hline
\multicolumn{8}{c}{47\,Tuc}      \\
x01    &   \       &   \    &   7.2   & 13.8   &  1.85  &   2.5 &   \     \\[-0.00pt]
x02    &   \       &   14   &   0.5   &  0.29  &  0.08  &   0.9 &   \     \\[-0.00pt]
x03    &   \       &   5.8  &   1.2   &  0.96  &  0.24  &   1.0 &   \     \\[-0.00pt]
x04    &   \       &   11   &   0.5   &  0.31  &  0.09  &   2.1 &   \     \\[-0.00pt]
x05    &   \       &   2.1  &   2.1   &  2.2   &  0.45  &   11  &   \     \\[-0.00pt]
x06    &   \       &   \    &   8.7   & 17.8   &  2.23  &   4.8 &   \     \\[-0.00pt]
x07    &   \       &   \    &   9.3   & 21.1   &  2.55  &   3.4 &   \     \\[-0.00pt]
x08    &   \       &   \    &  19.6   & 64.5   &  5.73  &   46\rlap{$^{(g)}$}&\ \\[-0.00pt]
x09    &   \       &   72   &   2.7   &  3.6   &  0.66  &   \   &   \     \\
    \hline
\multicolumn{8}{c}{NGC\,6388}      \\
o01    &   \       &   \    &   1.1   &  0.71  &  0.19  &   \   &   \     \\[-0.00pt]
o02    &   0.34    &   13   &   3.4   &  4.4   &  0.78  &   \   &   20.1  \\[-0.00pt]
o03    &   \       &   14   &   0.5   &  0.28  &  0.08  &   \   &   \     \\[-0.00pt]
o04    &   \       &   \    &   1.2   &  0.88  &  0.22  &   \   &   \     \\[-0.00pt]
o05    &   \       &   46   &   0.5   &  0.30  &  0.09  &   \   &   \     \\[-0.00pt]
o06    &   \       &   \    &   1.3   &  1.1   &  0.26  &   \   &   \     \\[-0.00pt]
o07    &   \       &   1.4  &   0.5   &  0.31  &  0.08  &   \   &   \     \\[-0.00pt]
o08    &   \       &   35   &   1.1   &  0.81  &  0.21  &   \   &   \     \\[-0.00pt]
o09    &   \       &   79   &   0.4   &  0.19  &  0.07  &   \   &   \     \\[-0.00pt]
o10    &   \       &   36   &   0.5   &  0.28  &  0.09  &   \   &   \     \\[-0.00pt]
o11    &   \       &   2.0  &   0.7   &  0.38  &  0.12  &   \   &   \     \\[-0.00pt]
o12    &   \       &   2.1  &   1.0   &  0.62  &  0.17  &   \   &   \     \\
x01    &   \       &   \    &   0.9   &  0.52  &  0.15  &   1.4 &   \     \\[-0.00pt]
x02    &   \       &   3.5  &   1.0   &  0.69  &  0.18  &   1.4 &   \     \\[-0.00pt]
x03    &   \       &   \    &   2.1   &  2.09  &  0.44  &   2.2 &   \     \\[-0.00pt]
x04    &   \       &   \    &   0.8   &  0.44  &  0.13  &   1.2 &   \     \\[-0.00pt]
x05    &   \       &   2.4  &   1.4   &  1.20  &  0.29  &   6.8 &   \     \\[-0.00pt]
x06    &   \       &   \    &   2.0   &  1.87  &  0.41  &   2.3 &   \     \\[-0.00pt]
    \hline
\multicolumn{8}{c}{M\,15}      \\
x01    &   0.03    &   0.4  &   0.09  &  0.04  &  0.01  &   3.3 &   0.9   \\[-0.00pt]
x02    &   0.06    &   6.4  &   0.23  &  0.12  &  0.03  &   4.3 &   2.0   \\
    \hline
\multicolumn{8}{c}{$\omega$\,Cen}         \\
x01    &   \       &   \    &   2.3   &  1.67  &  0.38  &   \   &   \     \\
    \hline
\multicolumn{8}{p{0.43\textwidth}}{\small Equivalent widths are determined using residuals from the model fits. Mass-loss rates determined using or quoted from: (a) our model (section 5); (b) Reimers (1975), (c) Schr\"oder \& Cuntz (2005); (d) Nieuwenhuijzen \& de Jager (1990); (e) OFFR02; (f) Gratton (1983), assuming a shell temperature of 8500 K and using the peak of the emission wing as $v_\infty$; (g) rate for 47\,Tuc\,x08 from Frogel \& Elias (1988), using similar methods to OFFR02. The errors on these values are large and discussed in the text. \normalsize} \\
    \hline
\end{tabular}
\end{center}
\end{table}

\begin{figure*}[!btp]
\resizebox{\hsize}{!}{\includegraphics[angle=270]{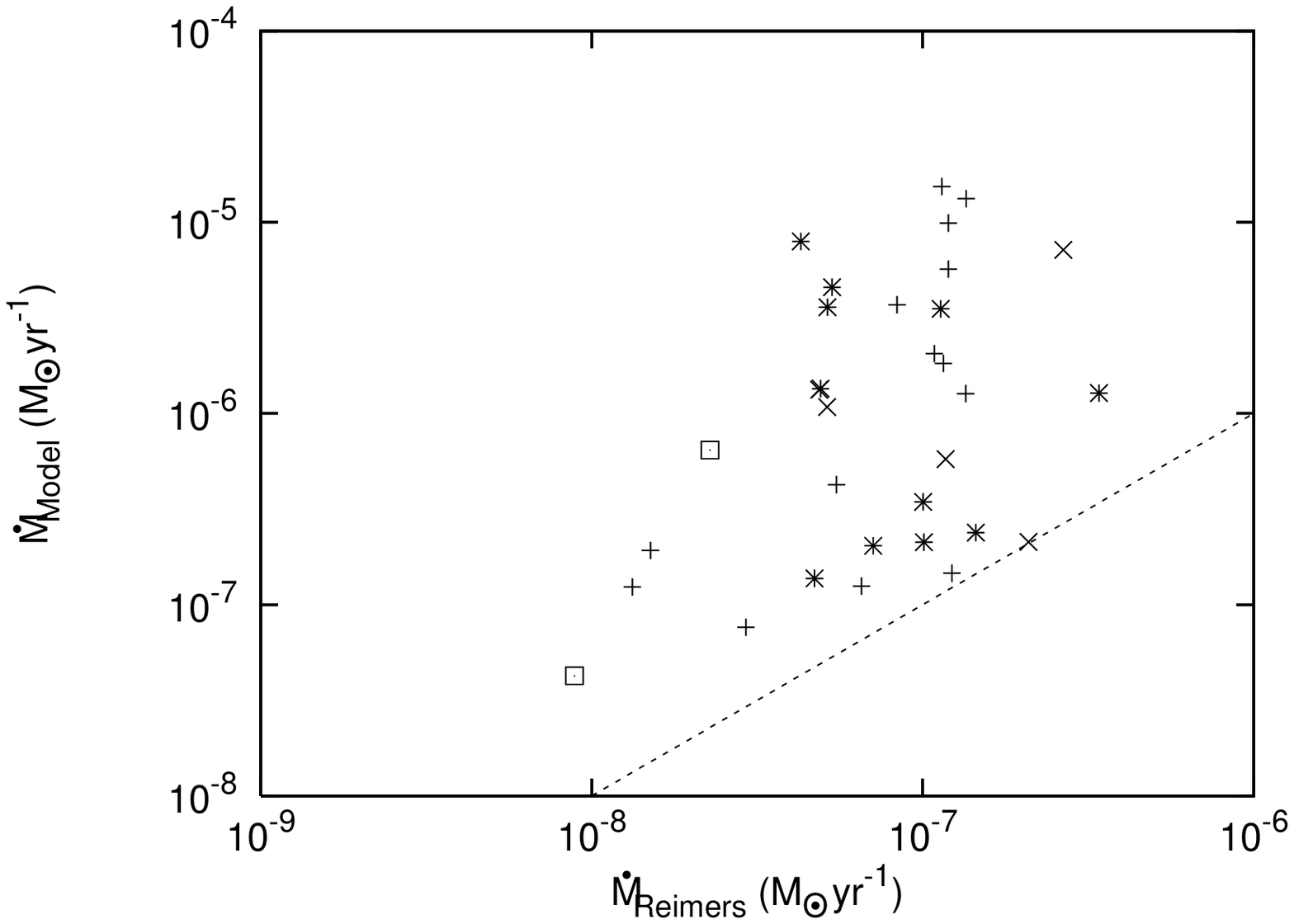}
                      \includegraphics[angle=270]{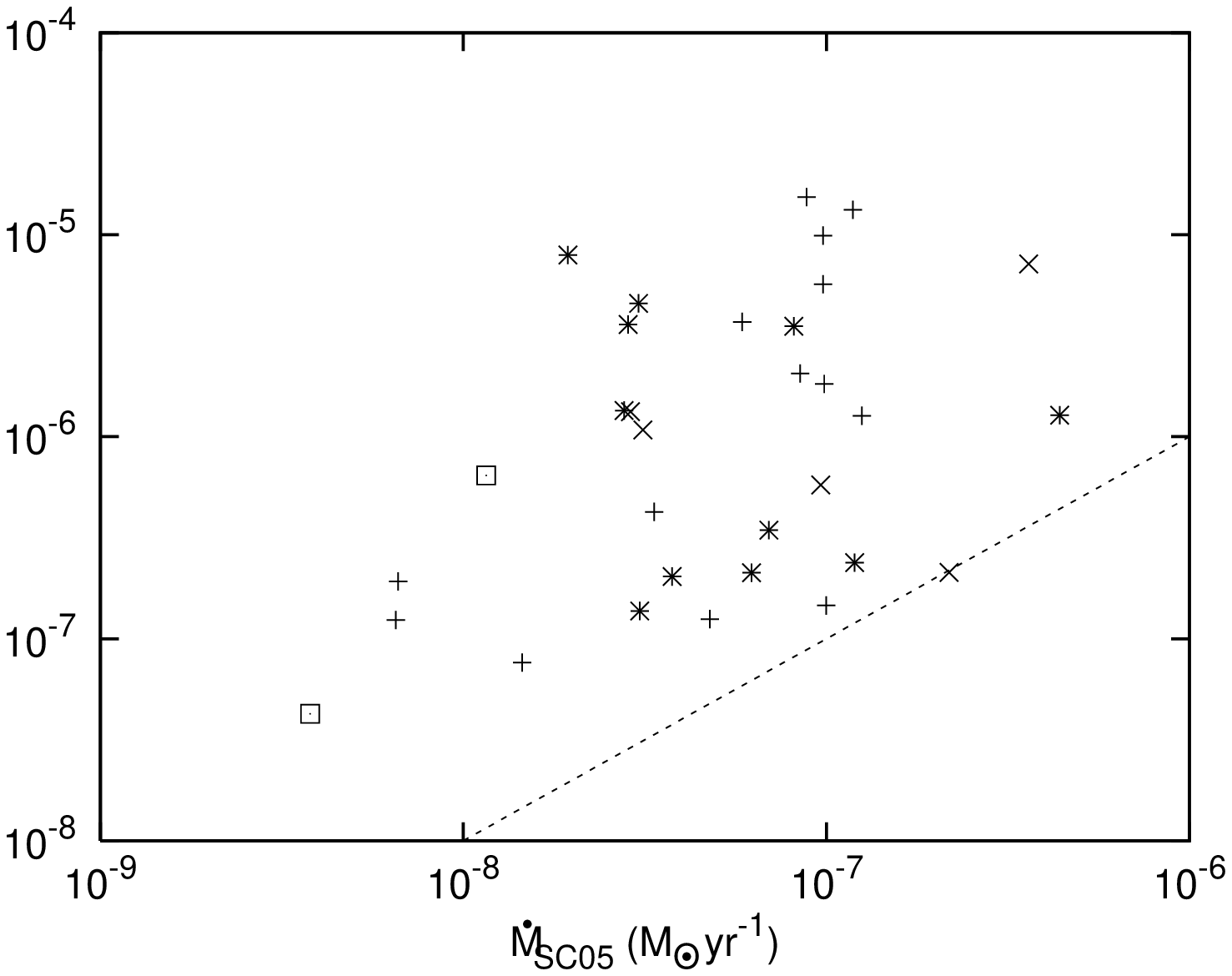}}
\resizebox{\hsize}{!}{\includegraphics[angle=270]{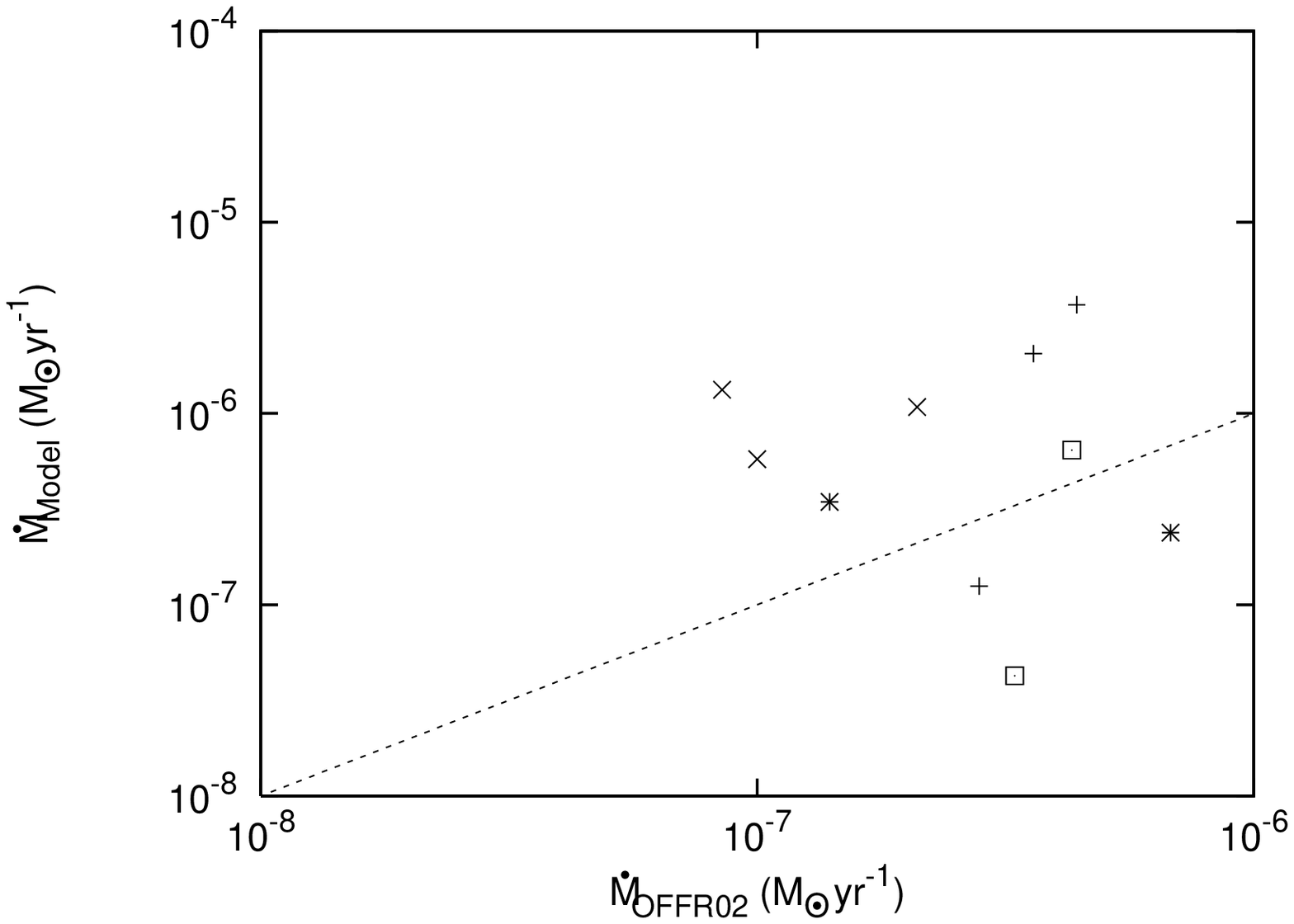}
                      \includegraphics[angle=270]{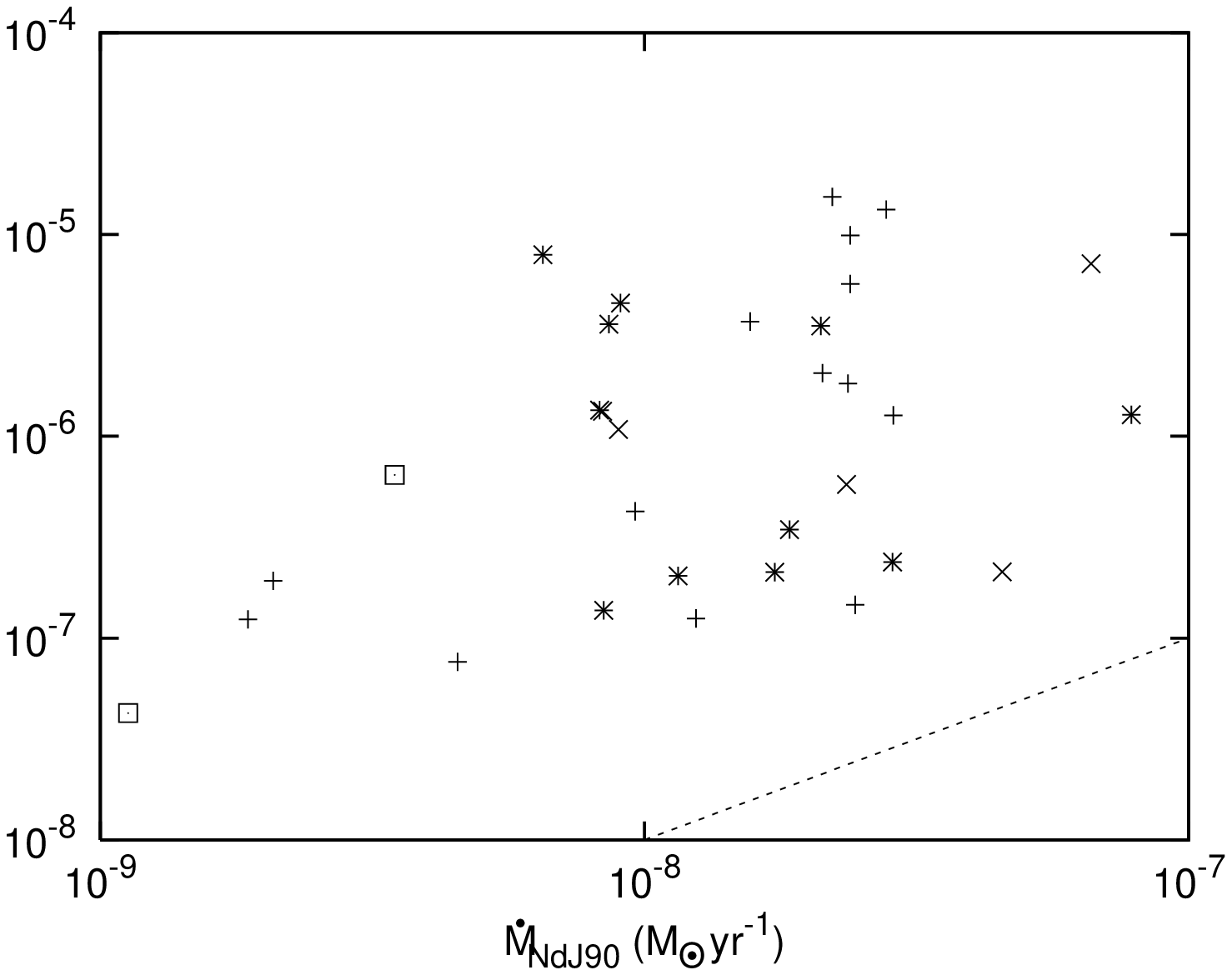}}
\caption[]{Comparison of estimated mass-loss rates from our modelling, Reimers (1975), OFFR02, SC05 and NdJ90. Key: NGC\,362: plus-signs, 47\,Tuc: crosses, NGC\,6388:  asterisks, M\,15: open squares. The dashed lines indicate a one-to-one correlation. Our mass-loss rates are very weakly correlated with, but significantly higher than those predicted by Reimers, SC05 and NdJ90, but in reasonable agreement with those observationally determined by OFFR02, which are determined through measured dust masses and dust temperatures.}
\label{MdotsFig}
\end{figure*}

Figure \ref{MdotsFig} shows a comparison between the first five methods in Table \ref{MassLossTable}. Although they weakly follow the predicted correlation with stellar parameters, all our targets clearly lie above the mass-loss rate predicted by all three semi-empirical relations. However, our mass-loss rates agree well with those determined by OFFR02 -- the observed scatter is concordant with our expected errors. As the OFFR02 results are derived independently from ours, it seems probable that our mass-loss rates -- at least for the IR-excessive stars -- are roughly correct.

The reason behind the lack of agreement with the semi-empirical relations is unclear, but could be influenced by several factors. Though Reimers' formula focussed on giants, the mass-loss rates for all three semi-empirical relations are mainly derived from, on average, more massive, more luminous and hotter stars than we see here. The Reimers and SC05 models also do not make allowances for winds influenced by dust or pulsation, which are undoubtedly important in some of our sample. It has also been suggested that the NdJ90 models use an unphysical prescription based on a mix of mass-loss rates from dust-driven and non-dust-driven sources (Schr\"oder \& Cuntz 2007). Furthermore, the models make no account of individual variations between stars outside of the parameters they use, such as pulsation phase and composition; they are also calibrated using only galactic stars, and relations for stars at near-solar metallicities may not necessarily hold for low-metallicity environments such as these. Origlia et al.'s values, however, are calculated for each star individually using the {\sc dusty} model (Ivezi\'c et al.\ 1999). The systematically higher mass-loss rates that we find with respect to the predictions from semi-empirical relations cannot be accounted for merely by a bias in the target selection procedure (Section 2), and it suggests instead that the efficiency of mass loss for low-mass, metal-poor red giants at the tip of the RGB and AGB is underestimated by the semi-empirical relations.

\begin{figure}[!btp]
\resizebox{\hsize}{!}{\includegraphics[angle=270]{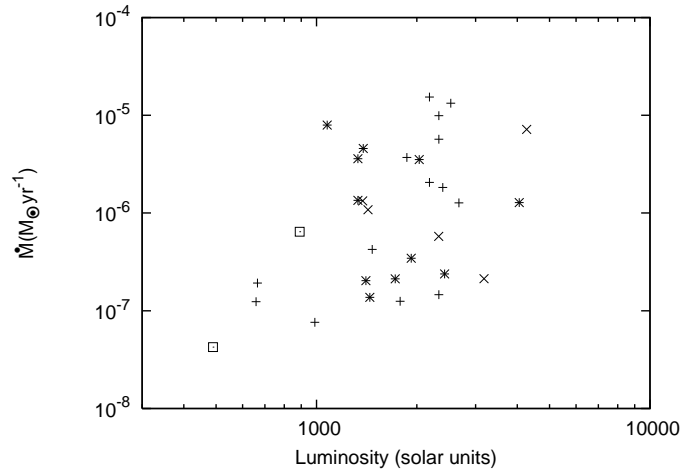}}
\caption[]{Variation of mass-loss rate with luminosity. Key is as in Figure \ref{MdotsFig}.}
\label{ModelLMFig}
\end{figure}

In exploring the correlation between our mass-loss rates and velocities and various other stellar parameters, we find little correlation among any of them. Perhaps the strongest correlation is between mass-loss rate and luminosity (Fig.\ \ref{ModelLMFig}), which shows a general increase in mass-loss with luminosity (hence evolution), in a similar fashion as is seen in more massive dust-enshrouded AGB stars in the Magellanic Clouds (e.g. van Loon et al.\ 2005). Note that the coolest and most luminous stars which show the strongest emission have not been modelled, but are expected to have mass-loss rates near the top of this distribution, which would further enhance this correlation.

\begin{figure}[!btp]
\resizebox{\hsize}{!}{\includegraphics[angle=270]{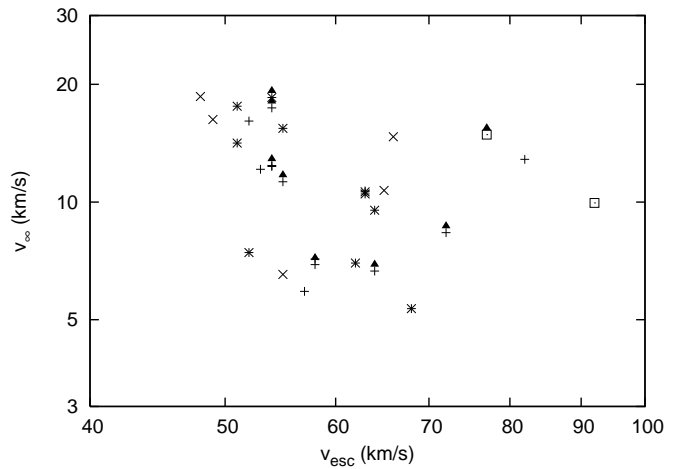}}
\caption[]{Variation of wind velocity with escape velocity. Key is as in Figure \ref{MdotsFig}. Attached triangles denote lower limits.}
\label{ModelVeViFig}
\end{figure}

\begin{figure}[!btp]
\resizebox{\hsize}{!}{\includegraphics[angle=270]{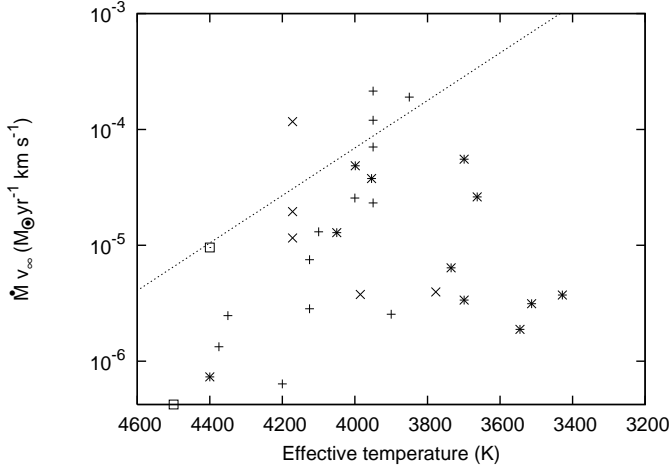}}
\caption[]{Variation of wind momentum with stellar effective temperature. Key is as in Figure \ref{MdotsFig}. The dotted line represents the fit by Judge \& Stencel (1991) for field giants.}
\label{ModelTMVFig}
\end{figure}

We also present the relationships between modelled wind velocity and computed escape velocity (Fig.\ \ref{ModelVeViFig}) and wind momentum versus effective temperature (Fig.\ \ref{ModelTMVFig}) after JS91. In the first instance, like JS91, we do not find any correlation between the two parameters, but we do find similiar average velocities around 10 km s$^{-1}$. Our wind momenta follow roughly the relation they suggest, but the momentum we derive appears to be several times lower for a given temperature. We suggest that this is due to the fact that our stars are less massive (likely $\sim$0.5--0.8 M$_{\odot}$, depending mostly on mass-loss history), wheras some of their sample could be up to 4 M$_{\odot}$. We do not find that wind velocity decreases with later spectral type (c.f.\ point four in JS91's conclusions). We also cannot support the claim that wind velocity increases with luminosity, as Cahn \& Wyatt (1978) suggest for field stars. It is also noteworthy that the velocities we find are significantly lower than the escape velocity of the star. Given our results are most sensitive to the material within the first few stellar radii, this suggests that the material \emph{must} receive an accelerating force long after it leaves the stellar surface. Once again, with their strong pulsations the coolest (unmodelled) stars are likely to have large wind momenta and thus contribute to the correlation observed. We do not find any correlation between metallicity and wind velocity, nor turbulent velocity (c.f. Gratton et al.\ 1984).

\begin{figure}[!btp]
\resizebox{\hsize}{!}{\includegraphics[angle=270]{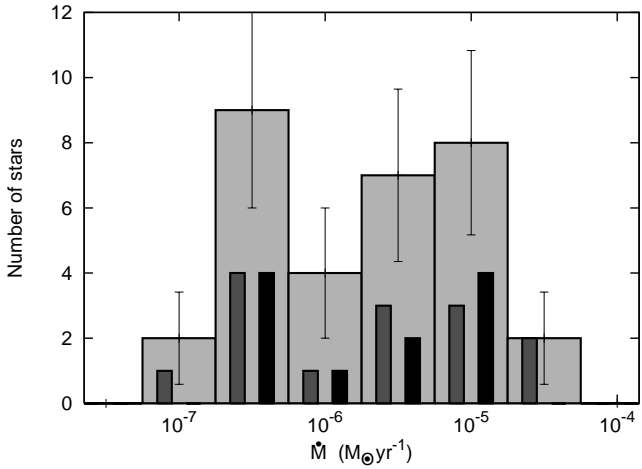}}
\caption[]{Histogram of mass-loss rate. The entire sample is represented in light grey, with the corresponding Poissonian errors; NGC\,362 is shown on the left of each bin in dark grey and NGC\,6388 is shown in black on the right of each bin. Note the bimodality seen both in each cluster and the sample as a whole.}
\label{ModelMdotHistFig}
\end{figure}

Both Figs.\ \ref{ModelLMFig} \& \ref{ModelVeViFig} show broad scatter, the former particularly towards higher luminosities. This scatter is greater than that expected from our errors, and could possibly represent variation of mass-loss rate with the phase of the pulsation cycle or from differences between the RGB and AGB. In both figures there are hints of a bimodality in both NGC\,362 and NGC\,6388 individually, and in the sample as a whole. A histogram in Figure \ref{ModelMdotHistFig} shows that the distribution is indeed slightly bimodal, avoiding the mass-loss region of $\sim$10$^{-6}$ M$_{\odot}$ yr$^{-1}$, though the strength of this bimodality depends sensitively on the size of the bins.

The cause of this bimodality is not clear. It may correspond to an `on' and an `off' phase of mass-loss as the star expands and contracts, or could be related to chromospheric instability. The lack of correlation with luminosity suggests that this is not a product of two different evolutionary phases, though considering there is only one star in the low-mass-loss region that is clearly above the RGB tip (several others could be below it within the errors), the bimodality may reflect a difference in mass-loss on the RGB and AGB. Were this the case, the Reimers and SC05 mass-loss laws would still hold for the weaker RGB winds, where dust and pulsation may not be important, though they would not apply to the stronger, dusty and/or pulsation-driven winds of the AGB stars.

The caveat to our results is that it becomes difficult to explain the high mass-loss rates derived for these temperatures unless all the stars for which we have calculated a mass-loss rate are in a brief phase of particularly high mass loss. Using Cioni et al.'s models, we find that an (initially) 0.8 M$_\odot$ star loses 0.13 M$_\odot$ on the portion of the AGB above the luminosity of the RGB tip ($\approx 1900$--2200 L$_{\odot}$, depending on metallicity) over a period of around 2.5 Myr, placing a limit on the \emph{average} mass-loss rate of 5.2 $\times 10^{-7}$ M$_{\odot}$ yr$^{-1}$ during that phase. The average value we find for these stars from modelling is 4.2 $\times 10^{-6}$ M$_{\odot}$ yr$^{-1}$, and may be larger, depending on the mass-loss rates of the most luminous stars that were not modelled. We also note that the highest mass-loss rates we find ($\gtrsim 4 \times 10^{-6}$ M$_{\odot}$ yr$^{-1}$) are among stars without IR excess or observed pulsation (except for 47\,Tuc\,x09), but also without strong evidence of chromospheric activity (line wing emission) in all but two of the less extreme cases (NGC\,362\,o07 and x01). These stars are mostly around spectra types K1--K3 and may have derived temperatures which are too low for their spectral type (see Fig.\ \ref{TempsFKFig}); the later type stars this applies to are all in the metal-rich NGC\,6388 and the H$\alpha$ line may be significantly affected by molecular lines that were not correctly modelled.

If we assume the presence of a chromosphere, we can reduce our derived mass-loss rates to lie more in line with theory. However, this may mean that the results given by OFFR02 are also too high. This could be due to incorrect assumptions on their part, particularly in the assumed gas-to-dust ratio (although that would require near-solar values even in metal-poor stars). We find no significant deviation between our mass-loss rates and those of Origlia et al.\ among the clusters, suggesting that their scaling of gas-to-dust with metallicity may be correct.

\subsection{Shell temperatures and chromospheres}

In order to further explore the properties of the chromosphere and outflow, we also compute mass-loss rates from the H$\alpha$ emission wings using the formula modified from Cohen (1976):
\begin{equation}
	\dot{M} = \frac{4}{3} \eta v_\infty R_{\ast} (R_s W_\lambda)^{0.5} e^{-1.1 / T_4},
\label{MdotG83}
\end{equation}
(in M$_{\odot}$ yr$^{-1}$) where $R_{\rm s}$ is the radius of the shell in stellar radii (assumed to be around $2 R_{\ast}$, though this value is not critical when compared to the uncertainty in the other parameters), $W_\lambda$ is the equivalent width (in \AA) of the emission line (see Table \ref{MassLossTable}), and $T_4$ is the temperature of the shell in units of $10^4$ K. The constant $\eta$ has been given as $6.6 \times 10^{-10}$ by Gratton (1983), which we adopt here; and as $2.4 \times 10^{-11}$ by Cacciari \& Freeman (1983) (both derived using G \& K stars). The origin of the difference between these two values is unclear. The factor $\frac{4}{3}$ arises from the conversion from hydrogen mass-loss rate to total mass-loss rate (ignoring molecular hydrogen).

\begin{figure}[!btp]
\resizebox{\hsize}{!}{\includegraphics[angle=270]{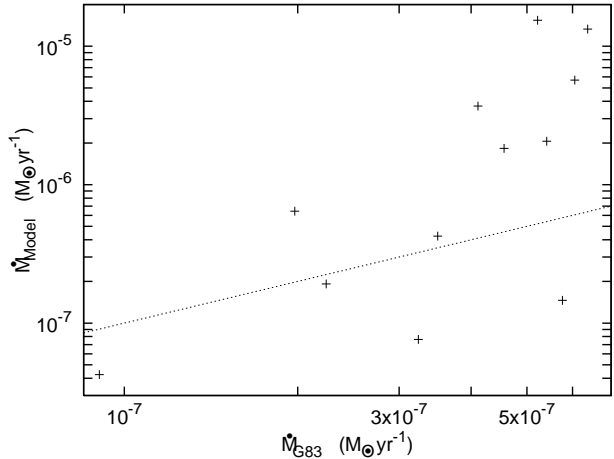}}
\caption[]{Comparison of mass-loss rates between the model of Gratton (1983) and our model. The dotted line indicates a one-to-one correlation. The expansion velocities used in the calculation of the Gratton mass-loss rate are taken from the emission peak, in keeping with their analysis.}
\label{ModelMGMMFig}
\end{figure}

It should be noted that Cohen's model (used by Gratton and Cacciari \& Freeman) bases the outflow velocity on the peak of the emission wing. These velocities are typically a factor $\sim$4 larger than the velocities we obtain from modelling the absorption profile. We follow their procedure and estimate the velocity from the emission wings, which are listed in Table \ref{VelsTable}. Gratton (1983) calculates $T_4$ as:
\begin{equation}
	T_4 = 10^{-4} \frac{T_{\ast}}
	      {1 + (T_{\ast} / 2.2 \times 10^4) \ln [R_{\rm s}^2 / W_\lambda]},
\label{T4G83}
\end{equation}
produces temperatures of typically around 2400--2550 K for all stars due to the terms in the denominator. It is difficult to see how a shell at this temperature could produce the substantial amount of emission seen. On the contrary, exploration of our model suggests that the shell temperatures must be $\gtrsim$ 6000 K in order to produce the observed line profiles, though we cannot determine this accurately using only the physics we have incorporated into the model. This latter figure would be in better agreement with other predictions, such as those of Falceta-Gon\c{c}alves et al.\ (2006). We assume a shell temperature of 8500 K (following Dupree et al.\ 1984). The mass-loss rates thus computed with Eq.\ (\ref{MdotG83}) are listed in Table \ref{MassLossTable} and plotted in Fig.\ \ref{ModelMGMMFig} where we compare them with the mass-loss rates from our model. There is a fair correlation and agreement between Gratton's and our models, except that our model suggests an order of magnitude higher rates for some of the more extreme stars.

\begin{table}[!Htbp]
\begin{center}
\caption{Comparison of wind velocities and escape velocities.}
\label{VelsTable}
\begin{tabular}{l@{\qquad\quad},,,@{\qquad\quad}cc}

\hline\hline
ID    & \multicolumn{5}{c}{Velocity (km s$^{-1}$)} \\[-2pt]
      & \multicolumn{3}{c}{$v_\infty$}&\multicolumn{2}{c}{$v_{\rm esc}$}\\
      & \multicolumn{3}{c}{$\overbrace{\qquad\qquad\qquad\qquad\qquad}$}
      & \multicolumn{2}{c}{$\overbrace{\qquad\qquad\qquad}$} \\[-3pt]
      & \multicolumn{1}{c}{\ \ Core\ \ }&\multicolumn{1}{c}{\ \ Model\ \ }&\multicolumn{1}{c}{\ \ Wing\ \ }& \multicolumn{1}{c}{\ R$_\ast$\ }&\multicolumn{1}{c}{\quad 3\,R$_\ast$}\\
      & \multicolumn{1}{c}{(a)}&\multicolumn{1}{c}{(b)}&\multicolumn{1}{c}{(c)}&\multicolumn{1}{c}{(d)}&\multicolumn{1}{c}{\quad(e)}\\
    \hline
\multicolumn{6}{c}{NGC\,362} \\
o02 &   1.9 &  8.4 & 43.5 & 72 & 41 \\
o03 &  -0.7 & 17.4 & 43   & 54 & 31 \\
o04 &   2.4 &  6.7 & 44.5 & 64 & 37 \\
o05a&   6.5 &  7.2 & 41   & 56 & 32 \\
o07 &   6.2 & 12.4 & 40.5 & 54 & 31 \\
o08 &   7.9 & 12.3 & 41   & 54 & 31 \\
o09 &   5.4 & 12.9 & 43   & 82 & 47 \\
x01 &   3.8 &  6.9 & 42.5 & 58 & 34 \\
x02b&  12.5 & 16.1 & 43.5 & 53 & 31 \\
x03 &   3.5 & 11.3 & 40   & 55 & 32 \\
    \hline
\multicolumn{6}{c}{NGC\,6388} \\
o02 &  -4.7 &  1.5 & 60   & 43 & 25 \\
    \hline
\multicolumn{6}{c}{M\,15} \\
x01 &   4.2 & 10.0 & 44.5 & 92 & 53 \\
x02 &  12.9 & 14.9 & 47   & 77 & 45 \\
    \hline
\multicolumn{6}{p{0.43\textwidth}}{\small Velocities: (a) blueshifted H$\alpha$ line core velocity; (b) $v_\infty$ from our model; (c) line wing emission velocity (averaged where both wings show emission);  (d) escape velocity at surface; (e) escape velocity at 3 R$_\ast$. The emission in NGC\,6388\,o02 may be due to pulsation, but it is included here for completeness. \normalsize} \\
    \hline
\end{tabular}
\end{center}
\end{table}

The velocities we list in Table \ref{VelsTable} are from the line wing emission peaks, for consistency with the previous works. They are identical within the $\sim$5--10\% error to those velocities found from measuring the peak of the residual spectrum after subtraction of our model. Interestingly, these velocities are comparable to the escape velocity found at around $\sim$3 R$_\ast$. Assuming the shell is at roughly this radius from the star (both Gratton and Cacciari \& Freeman assume 2 R$_\ast$), this weakens the need for a mechanism to drive the wind beyond this radius, thus allowing a situation where mass-loss can occur solely through the chromosphere. In the case of NGC\,6388\,o02, pulsation may already be strong enough to eject matter from the surface.

Additionally, an anti-correlation exists between escape velocity and wind velocity as measured by our model (Fig.\ \ref{ModelVeViFig}); though it does not exist in the sample as a whole, the anti-correlation may be weakly seen on a per-cluster basis. This would suggest that gravity inhibits the wind acceleration.

\subsection{Chromospheric disruption and the metallicity dependence of mass loss}

Our observations suggest that, as luminosity increases and the stellar surface cools, the dominance of chromospheric activity is replaced by radial pulsation. Unless by coincidence the chromosphere switches off at the point where significant pulsation begins, it seems likely that pulsation may play a more active r\^ole in diminishing the chromospheric control on atmospheric behaviour. Either pulsation disrupts the mechanism responsible for heating the chromosphere, or the pulsation-induced shocks overrun the chromosphere and displace the hot gas to further out in the wind where the shocks dissipate, turning the expanding column in front of the star (which provides absorption and hence blue asymmetry in the profile) into an emitting column of material. This leads to a weakening of the line core and the onset of the type of profiles seen in NGC\,362\,o01 and NGC\,6388\,x03, and then those seen in 47\,Tuc\,x08 and $\omega$\,Cen\,x01. 

The transition from chromospheric activity to pulsation-related activity is not absolute, and the wide variety of profiles may represent either pulsation or a chromosphere dominating activity, depending on the phase of the pulsation cycle. This transition appears to happen around spectral type K3--K5.5 in NGC\,362, with pulsation-related shocking becoming noticable at around spectral types M1--M2 in NGC\,6388. This is much earlier than the $\sim$M5 spectral type for field stars found by JS91, which is highly suggestive of a metallicity relation among the three. This corresponds to temperatures of $\sim$3500 K in the field stars, $\sim$3700 K in NGC\,6388 and $\sim$3800--4000 K in NGC\,362 (see Fig.\ \ref{TempsFKFig}). As can be seen from Figs.\ \ref{HalphaMinusKuruczFig} \& \ref{HalphaMinusKuruczFig2}, it is not clear from our sample whether the transition is more dependent on temperature or luminosity, though these temperatures would all roughly correspond to the RGB tip (see Fig.\ \ref{UVESHRD}). It is possible that our lack of differentiation between the RGB and AGB could be influencing our conclusions, though pulsation-induced shocks are only visible above the RGB tip. Interestingly, these temperatures also mark the divergence of the temperature vs.\ spectral type relations of Levesque et al.\ (Fig. \ref{TempsFKFig}), though it is not obvious why the two phenomena should be linked.

The fractional occurence of H$\alpha$ emission was investigated by Gratton et al.\ (1984), who find that 70\% of stars above 500 L$_{\odot}$ (with no metallicity variation) and $\sim$80\% of stars above 2000 L$_{\odot}$ show H$\alpha$ emission in some form, though they did not differentiate between chromospheric-like emission and pulsation-related emission. Our data show 37\% and 42\%, respectively, for the same statistics. We have only one star (M\,15\,x01) below 500 L$_{\odot}$ and, of the remainder, ten (22\%) are presumed to have chromosphere-induced emission and seven (15\%) exhibit pulsation-induced emission (on the assumption NGC\,6388\,o02 exhibits pulsation-induced emission), with a strong metallicity bias towards the presence of emission in the two most metal-poor clusters. Clearly the results are disparate. This may be a result of a higher average metallicity of our sample (Gratton et al.'s sample averages [Fe/H] $= -1.35$ whereas ours averages [Fe/H] $= -0.91$), which in turn could affect the visibility of H$\alpha$ emission through masking by stronger atomic or molecular lines due to higher abundances and lower temperatures. Their sample only contains two stars from 47\,Tuc and none from NGC\,362 or 6388, so we cannot perform a comparative study on individual clusters.

The mass-loss rates appear to be no different for stars in metal-poor and metal-rich clusters with similar properties, (Fig.\ \ref{MdotsFig}) such as luminosity (Fig.\ \ref{ModelLMFig}). This correlates with the statement made by JS91, that mass-loss rate in this phase of evolution does not strongly depend on the driving process.


\section{Conclusions}

In this study, we have presented VLT/UVES data of a set of giant branch stars in globular clusters, which are among the highest resolution, high signal-to-noise spectra of their kind. We use these to characterise and quantify the outflow from their atmospheres.

We have used Kurucz's {\sc atlas9} models to determine stellar temperature and, from this, we have calculated basic stellar parameters, which appear typical for red and asymptotic giant branch stars.

Many of the stars we have investigated show clear emission in H$\alpha$, most spectacularly when strong pulsation is present, which is seen to occur at luminosities above the RGB tip. It is also noted that many of the stars show strongly blue-shifted absorption cores, suggesting bulk outflow from the stellar surface. This is also mirrored in some cases in the near-infrared calcium triplet line profiles.

In an effort to quantify this outflow, we have constructed a simple model for the warmer and more metal-poor stars. We have calculated terminal velocities for the winds which are of order 10 km s$^{-1}$ -- much lower than the escape velocity of the stars -- and mass-loss rates of $\approx$ 10$^{-7}$ to 10$^{-5}$ M$_\odot$ yr$^{-1}$, which lie well above theoretical expectations, but are consistent with the mass-loss rates derived from IR emission from circumstellar dust. These models also suggest that an emissive shell exists close to the stellar surface in many stars, with the wind probably being largely isothermal beyond it.  Outflow velocities of the emission are of order 40 km s$^{-1}$, and mass-loss rates derived from this emission are similar to the ones we obtain from modelling the absorption profile. In the most extreme cases we may have either over-estimated the mass-loss rate from our model, or under-estimated the temperature of the emitting shell. Mass-loss rates correlate weakly with luminosity, but stars showing strong IR excesses (linked with dust production) do not necessarily exhibit higher gas mass-loss rates. We find no correlation between mass-loss rate and metallicity.

We suggest that the emission and mass-loss in early-type ($\lsim$ K3--K5) giants are dominated by a warm chromospheric region, as suggested by some studies, though late-type ($\gsim$ M3) giants have mass loss dominated by pulsation. It seems likely that the spectral type of the transition between the two regimes is metallicity-dependent, occurring at later spectral types for higher metallicities. The outward velocity of the warm emissive shells associated with the chromospheres is similar to the escape velocity at their anticipated radius (2 to 3 R$_\ast$), allowing the chromosphere to be the sole driver of mass loss in these stars.

\begin{acknowledgements}
We thank the ESO staff at Paranal for their excellent support, Livia Origlia, Wolf-Rainer Hamann, John J. Keady and Peter Wood for valuable discusions, and Klaus-Peter Schr\"oder for his helpful comments as referee. Iain McDonald is supported by an STFC studentship.
\end{acknowledgements}


\end{document}